\documentclass[12pt,a4paper,twoside]{article}
\usepackage{amsmath}
\usepackage{amssymb}   
\usepackage{amsfonts}
\usepackage{theorem}
\usepackage{epsfig}
\usepackage{helvet}

\newcommand{\R}{{\mathord{\mathbb R}}}
\newcommand{\Z}{{\mathord{\mathbb Z}}}
\newcommand{\N}{{\mathord{\mathbb N}}}
\newcommand{\C}{{\mathord{\mathbb C}}}

\newcommand{\T}{{\mathord{\mathbb T}}}

\newcommand{\mH}{{\mathcal H}}
\newcommand{\mB}{{\mathcal B}}

\newcommand{\mL}{{\mathcal L}}
\newcommand{\mS}{{\mathcal S}}

\newcommand{\mE}{{\mathcal E}}
\newcommand{\mO}{{\mathcal O}}
\newcommand{\mQ}{{\mathcal Q}}

\newcommand{\rB}{{\rm B}}
\newcommand{\dd}{{\rm d}}
\newcommand{\e}{{\rm e}}
\newcommand{\raa}{{\rm aa}}
\newcommand{\rac}{{\rm ac}}
\newcommand{\rap}{{\rm ap}}
\newcommand{\rpa}{{\rm pa}}
\newcommand{\rpp}{{\rm pp}}
\newcommand{\rsc}{{\rm sc}}
\newcommand{\rt}{{\rm t}}
\newcommand{\ii}{{\rm i}}

\newcommand{\ran}{{\rm ran\,}}
\newcommand{\sign}{{\rm sign}}

\def\slim{\mathop{\rm s-lim}}
\def\wlim{\mathop{\rm w-lim}}
\newcommand{\hh}{{{\mathfrak h}}}

\newcommand{\ff}{{{\mathfrak f}}}

\newcommand{\fF}{{\mathfrak F}}
\newcommand{\fA}{{\mathfrak A}}

\newcommand{\pf}{{\rm pf}}

\newcommand{\ind}{{\rm ind}}
\newcommand{\eps}{{\varepsilon}}

\newcommand{\spec}{{\rm spec}}

\newcommand{\lN}{\ell^2(\N)}
\newcommand{\rP}{{\rm P}}
\newcommand{\kp}{\kappa}
\newcommand{\vi}{\varphi}
\newtheorem{thm}{Theorem}
\newtheorem{proposition}[thm]{Proposition}
\newtheorem{lemma}[thm]{Lemma}
\newtheorem{definition}[thm]{Definition}
{\theorembodyfont{\upshape} \newtheorem{remark}[thm]{\it Remark}}

\newtheorem{assumption}[thm]{Assumption}
\newtheorem{corollary}[thm]{Corollary}
\newcommand{\bd}{\begin{definition}}
\newcommand{\ed}{\end{definition}\vspace{0mm}}
\newcommand{\bes}{\begin{eqnarray*}}
\newcommand{\ees}{\end{eqnarray*}}
\newcommand{\be}{\begin{eqnarray}}
\newcommand{\ee}{\end{eqnarray}}
\newcommand{\bt}{\begin{thm}}
\newcommand{\et}{\end{thm}\vspace{0.5mm}}
\newcommand{\bc}{\begin{corollary}}
\newcommand{\ec}{\end{corollary}\vspace{0.5mm}}
\newcommand{\bl}{\begin{lemma}}
\newcommand{\el}{\end{lemma}\vspace{0.5mm}}
\newcommand{\bp}{\begin{proposition}}
\newcommand{\ep}{\end{proposition}\vspace{0.5mm}}
\newcommand{\br}{\begin{remark}}
\newcommand{\er}{\end{remark}}
\newcommand{\ba}{\begin{assumption}}
\newcommand{\ea}{\end{assumption}}
\newcommand{\bprf}{{\it Proof.}\, }
\newcommand{\eprf}{\hfill $\Box$ \vspace{5mm}}
\begin{document}
\pagestyle{myheadings}

\markboth{W. H. Aschbacher}{Broken translation invariance
in quasifree nonequilibrium correlations}

\title{Broken translation invariance in quasifree fermionic correlations out of equilibrium}

\author{Walter H. Aschbacher\footnote{walter.aschbacher@polytechnique.edu}
\\ \\
Ecole Polytechnique\\
Centre de Math\'ematiques Appliqu\'ees\\
UMR CNRS - 7641\\
91128 Palaiseau Cedex\\
France}

\date{}
\maketitle
\begin{abstract} 
Using the $C^\ast$ algebraic scattering approach to study quasifree fermionic systems out of equilibrium in quantum statistical mechanics, we construct the nonequilibrium steady state in the isotropic XY chain whose translation invariance has been broken
by a local magnetization and analyze the asymptotic behavior of the expectation value for a class of spatial correlation observables in this
state.
The effect of the breaking of translation invariance is twofold. Mathematically, the finite rank perturbation not only regularizes the scalar symbol of the invertible Toeplitz operator generating the leading order exponential
decay but also gives rise to an additional trace class Hankel 
operator in the correlation determinant.
Physically, in its decay rate, the nonequilibrium steady state exhibits
a left mover--right mover structure affected by the scattering at
the impurity.
\end{abstract}
\section{Introduction}

In the mathematical study of open quantum systems, the role played by 
quasifree fermionic systems is an important one. Within the framework 
of algebraic quantum statistical mechanics, they not only allow for a
powerful description by means of scattering theory on the one-particle
Hilbert space over which the fermionic algebra of observables is built,
being thus ideally suited for rigorous analysis on many levels, but they
also represent a class of systems which are indeed realized in nature, see, for example, Culvahouse {\it et al.} \cite{CSP}, D'Iorio {\it et al.} \cite{DAT},  and Sologubenko 
{\it et al.} \cite{SGOVR}.
 A special instance of this class is the finite
XY spin chain introduced by Lieb {\it et al.} \cite{LSM} and extended 
to the infinite two-sided discrete line by Araki \cite{A3} in the framework of 
$C^\ast$-dynamical systems. As a matter of fact, this spin model can be 
mapped, in some precise sense, onto a gas of free fermions with the help of the Araki-Jordan-Wigner transformation. In order to study the effect of the breaking of translation invariance in this system, we choose the 
physically interesting and computationally convenient emptiness formation correlation observable. The so-called emptiness formation probability (EFP), {\it i.e.} the expectation value of this observable in a given state, describes, in the spin picture,  the probability that all spins in a string of a given length point downwards. However, we would like to underline that the analysis is not limited to this observable but can rather be carried out for a broad class of spatial correlations.

The asymptotic behavior of the EFP in the XY chain for large string length has already been analyzed for the cases where the state  is a ground state or a thermal equilibrium state at positive 
temperature. In both cases, the EFP can be written as the determinant of the  section of a Toeplitz operator with scalar symbol.  Since the higher order asymptotics of a Toeplitz determinant is highly sensitive to the
regularity of the symbol of the Toeplitz operator, the asymptotic behavior of the ground state EFP is qualitatively different in the 
so-called  critical and noncritical regimes corresponding to
certain  values of the anisotropy and the exterior magnetic field of the XY chain, {\it i.e.}, in \eqref{H-XY} below, the parameters 
$\gamma$ and $\lambda$, respectively. It has been found that the EFP decays like a Gaussian in one of the critical regimes (with some  additional explicit numerical prefactor and some power law prefactor), see Shiroishi {\it et al.} \cite{STN} and references therein. 
In a second critical regime and in all noncritical regimes, the EFP decays exponentially (in contrast to the noncritical regimes, there is an additional power law prefactor in the second critical regime whose exponent differs from the one in the first critical regime), see Abanov and Franchini \cite{AF, FA}.
These results have been derived by using powerful theorems of 
Szeg\H o, Widom, and Fisher-Hartwig, and the yet unproven Basor-Tracy conjecture and some of its extensions, see Widom \cite{W} and
 B\"ottcher and Silbermann  \cite{BS1, BS2}. Furthermore, in thermal equilibrium at positive temperature,  the EFP can
again be shown to decay exponentially by using a theorem of Szeg\H o, 
see, for example,  Shiroishi {\it et al.}  \cite{STN} and  Franchini and Abanov \cite{FA}.

In contrast, out of equilibrium, the situation is more subtle.
The typical open system consists of a confined sample which is coupled to extended ideal reservoirs at different temperatures. Using this paradigm, a translation invariant nonequilibrium steady state (NESS) has been constructed  in Aschbacher and Pillet \cite{AP} for the XY chain using the scattering
approach  to algebraic quantum statistical mechanics developed by Ruelle 
\cite{R1} (for $\gamma=\lambda=0$, this NESS has also been found by
Araki and Ho \cite{AH} using a different method; moreover, 
using the latter approach, the magnetization profile at intermediate but large times has been studied by Ogata \cite{O}).
In this NESS, the EFP can still be recast into the form of a Toeplitz determinant, but now, the symbol is, in general, no longer scalar
and regular. Due to the lack of control of
higher order determinant asymptotics in  Toeplitz theory with
nontrivial irregular block symbols, we started off by studying  bounds on the leading asymptotic order for a class of general block Toeplitz
determinants in Aschbacher \cite{1}.  There, it turned out that suitable basic spectral information on the density of the state is sufficient to derive a bound on the rate of the exponential decay of the EFP in general translation invariant fermionic quasifree states. 
This bound proved to be exact not only for the decay rates of the ground states and the equilibrium states at positive temperature treated in  Abanov and Franchini \cite{AF, FA} and Shiroishi {\it et al.} 
\cite{STN}  but also for the translation invariant NESS in the isotropic XY chain analyzed in Aschbacher \cite{3}.

In the present paper, new results are obtained for the asymptotic behavior of a class of spatial correlations, and in particular, for the asymptotic behavior of the EFP, in a NESS of the isotropic XY chain  whose translation invariance has been broken by a local magnetization in the form of a finite rank perturbation. Although such a spatial correlation can again be transformed from its initial Paffian form into a scalar Toeplitz determinant, the effect of the breaking of translation invariance manifests itself in a regularization of the Toeplitz symbol
and the appearance of an additional Hankel operator whose symbol
is smooth. Hence, due to Peller's theorem, this operator is of trace class, and the spatial asymptotics is  governed by an exponential decay due to the invertibility of the Toeplitz operator at nonvanishing
temperature. Moreover, the decay rate, determined by the Toeplitz symbol,
exhibits the underlying left mover--right mover structure 
  affected by the scattering at the impurity (see also Aschbacher \cite{2} and Aschbacher and Barbaroux  \cite{AB} for the left mover--right mover structure of the NESS expectation of several other types of correlation observables).

The paper is organized as follows. In Section \ref{sec:setting},  we set the stage for the nonequilibrium XY chain with impurity, construct its NESS, and derive the basic expression for the NESS EFP.  Section 
\ref{sec:asymptotics} then contains the asymptotic analysis of the NESS
EFP. Several ingredients of the proofs have been transferred to
the appendix, as, for example, the construction of the wave operators
by means of stationary scattering theory or the summary of the spectral properties of the so-called magnetic Hamiltonian.
\section{Nonequilibrium setting}
\label{sec:setting} 
In this section, we will shortly summarize the setting for the system
out of equilibrium used in Aschbacher and Pillet \cite{AP}. In contradistinction
to the presentation there, we skip the formulation of the two-sided XY chain
as a spin system and rather focus directly on the underlying
$C^\ast$-dynamical system structure in terms of Bogoliubov automorphisms on a selfdual CAR algebra as in Araki \cite{A3}.
A $C^\ast$-dynamical system is a pair $(\fA,\tau)$, where $\fA$ is a $C^\ast$ algebra and $\R\ni t\mapsto \tau^t\in {\rm Aut}(\fA)$ a strongly
continuous group of $\ast$-automorphism of $\fA$. For more information on the algebraic approach to open quantum systems, see, for example, 
Aschbacher {\it et al.} \cite{AJPP1}.

\begin{figure}
\setlength{\unitlength}{7mm}
\begin{picture}(-1,1)
\put(3,0){\line(1,0){4}}
\put(8,0){\line(1,0){6}}
\put(15,0){\line(1,0){4}}
\multiput(4,0)(1,0){4}{\circle*{0.2}}
\multiput(8,0)(1,0){7}{\circle*{0.2}}
\put(11,0){\circle*{0.3}}
\multiput(15,0)(1,0){4}{\circle*{0.2}}
\put(13.7,-0.7){$N$}
\put(7.3,-0.7){$-N$}
\put(18,0){\vector(1,0){1}}
\put(4,0){\vector(-1,0){1}}
\put(10.7,-1){$\Z_\mS$}
\put(16.5,-1){$\Z_R$}
\put(5.5,-1){$\Z_L$}
\put(11,0.1){\vector(0,1){1}}
\put(11.2,0.5){$\kappa$}
\end{picture}
\vspace{6mm}
\caption{The nonequilibrium setting for the XY chain.}
\label{fig:nonequilibrium}
\end{figure}
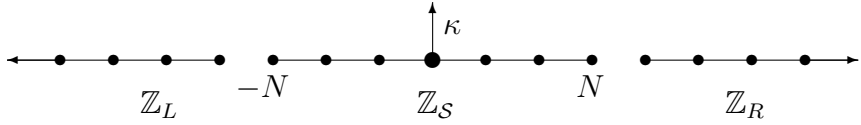

For some given $N\in\N\cup\{0\}$, the nonequilibrium configuration is set up by cutting the finite piece
\be
\Z_\mS
:=\{x\in\Z\,|\, -N\le x\le N\}
\ee
out of the two-sided discrete line $\Z$.  This piece will play the role of the confined sample whereas the remaining parts,
\be
\Z_L
&:=&\{x\in\Z\,|\, x\le -(N+1)\},\\
\Z_R
&:=&\{x\in\Z\,|\, x\ge N+1\},
\ee 
will act as infinitely extended thermal reservoirs, eventually carrying different temperatures, see Figure \ref{fig:nonequilibrium}.

The observables of the system are specified by the following selfdual
CAR algebra over the wave functions on the chain.
\bd[Observables]
\label{def:obs}
Let $\fF(\hh)$ denote the fermionic Fock space over the one-particle Hilbert space of wave functions on the discrete line,
\be
\hh
:=\ell^2(\Z).
\ee
With the help of the creation
and annihilation operators $a^\ast(f), a(f)\in \mL(\fF(\hh))$ with
$f\in\hh$ (where $\mL(\mH)$ denotes the bounded linear operators on the 
Hilbert space $\mH$), the complex linear mapping 
$B: \hh^{\oplus 2}\to \mL(\fF(\hh))$ is defined, for 
$F:=[f_1,f_2]\in  \hh^{\oplus 2}$, by
\be
\label{B}
B(F)
:=a^\ast(f_1)+a(\bar f_2).
\ee
The observables are described by the selfdual CAR algebra over 
$\hh^{\oplus 2}$ with antiunitary involution $J$ generated by the operators 
$B(F)\in\mL(\fF(\hh))$ for all $F\in\hh^{\oplus 2}$, {\it i.e.} we have,
for all $F,G\in \hh^{\oplus 2}$,
\be
\{B^\ast(F),B(G)\}
&=& (F,G),\\
B^\ast(F)
&=&B(JF),
\ee
where $JF:=[\bar f_2,\bar f_1]$
for all $F:=[f_1,f_2]\in  \hh^{\oplus 2}$, the anticommutator of 
$A,B\in\mL(\mH)$ is $\{A,B\}:=AB+BA$, and the scalar product in 
$\hh^{\oplus 2}$ is written as the one in $\hh$.
We denote this algebra by  
$\fA:=\fA(\hh^{\oplus 2},J)$.
\ed
\br
The concept of selfdual CAR algebras has been introduced and developed
in  Araki \cite{A1, A2}. Here, it is just a convenient way of working
with the linear combination \eqref{B}. Also in view of future
generalizations of the present paper, for example to the case of the truly anisotropic XY
chain and other classes of correlations, we will stick to this notation in
the present context.
\er

We next specify the Bogoliubov $\ast$-automorphisms on the selfdual 
CAR algebra which describe the time evolutions used for the  construction
of the NESS.
\bd[Dynamics]
\label{def:dynamics}
Let the coupling strength be $\kp>0$, and let $u\in\mL(\hh)$ be the translation given by 
$(uf)(x):=f(x-1)$ for all $f\in\hh$ and all $x\in\Z$. The XY, the decoupled, and the magnetic one-particle Hamiltonians 
$h, h_0, h_\rB\in\mL(\hh)$, respectively,  are defined by
\be
\label{h}
h
&:=&{\rm Re}(u),\\
\label{h0}
h_0
&:=& h-(v_L+v_R),\\
h_\rB
&:=& h +\kp v,
\ee
where the decoupling operators  $v_L,v_R\in \mL^0(\hh)$ 
(with $\mL^0(\mH)$ the finite rank operators on $\mH$)
and the operator $v\in\mL^0(\hh)$ which breaks translation invariance have the form
\be
v_L
&:=&{\rm Re}\big(u^{-(N+1)}p_0u^N\big),\\
v_R
&:=&{\rm Re}\big(u^Np_0u^{-(N+1)}\big),\\
v
&:=&p_0.
\ee
Here, the projection $p_0\in\mL^0(\hh)$ is given by 
$p_0 :=(\delta_0,\cdot\,) \delta_0$, where
$\delta_x\in\hh$ for $x\in\Z$ denotes the Kronecker function (moreover,
the real part of $A\in\mL(\mH)$ is given by ${\rm Re}(A):=(A+A^\ast)/2$).
For all $t\in\R$, the XY, the decoupled, and the magnetic time evolutions are the Bogoliubov $\ast$-automorphisms $\tau^t, \tau_0^t,
\tau_\rB^t \in {\rm Aut}(\fA)$ defined on the generators $B(F)\in\fA$ with $F\in \hh^{\oplus 2}$ by
\be
\tau^t(B(F))
&:=& B(\e^{\ii t H }F),\\
\tau^t_0(B(F))
&:=& B(\e^{\ii t H_0}F),\\
\tau^t_\rB(B(F))
&:=& B(\e^{\ii t H_\rB}F),
\ee
where we set $H:=h\oplus -h$, $H_0:=h_0\oplus -h_0$, and 
$H_\rB:=h_\rB\oplus -h_\rB$.
\ed
\br
For the sake of an easy exposition, we restrict the analysis
 to the case $\kp>0$, the case $\kp<0$ being strictly analogous.
\er
\br
The magnetic Hamiltonian $H_\rB\in\mL(\hh^{\oplus 2})$ breaks 
translation invariance in the sense that the commutator 
$[H_\rB,u\oplus u]=[h_\rB,u]\oplus - [h_\rB,u]$ is nonvanishing
(where $[A,B]:=AB-BA$ is the commutator of $A,B\in\mL(\mH)$), 
{\it i.e.}, for all $f\in\hh$, it holds
\be
[h_\rB,u]f
=\kp (f(-1)\delta_0-f(0)\delta_1).
\ee
\er
\br
\label{rem:JHB}
Since $H_\rB\in\mL(\hh^{\oplus 2})$ anticommutes with the antiunitary involution $J$,
the magnetic Hamiltonian $H_\rB$ generates a Bogoliubov transformation in the sense of Araki 
\cite{A1, A2}, {\it i.e.} that, 
for all $t\in\R$, we have
\be
\label{JHB}
[{\rm e}^{{\rm i}t H_\rB},J]=0.
\ee
The same also holds for the XY and the decoupled Hamiltonian 
$H, H_0\in\mL(\hh^{\oplus 2})$.
\er
\br
\label{rem:XY}
As mentioned at the beginning of this section, this model has its origin
in the XY spin chain whose formal Hamiltonian is given by 
\be
\label{H-XY}
H=-\frac{1}{4}\sum_{x\in\Z}\left\{(1+\gamma)\,\sigma_1^{(x)}\sigma_1^{(x+1)}+(1-\gamma)\,\sigma_2^{(x)}\sigma_2^{(x+1)}
+2\lambda  \, \sigma_3^{(x)}\right\},
\ee
where $\gamma\in(-1,1)$ denotes the anisotropy, $\lambda\in\R$
the external magnetic field, and the Pauli basis
of $\C^{2\times 2}$ reads
\be
\label{Pauli} 
\sigma_0
=\left[\begin{array}{cc}
1 & 0
\\ 0&1
\end{array}\right],\quad 
\sigma_1
=\left[\begin{array}{cc}
0 & 1\\
1& 0
\end{array}\right],\quad 
\sigma_2
=\left[\begin{array}{cc}
0 &-\ii\\ 
\ii & 0
\end{array}\right],\quad
\sigma_3
=\left[\begin{array}{cc}
1 & 0
\\ 0& -1
\end{array}\right].
\ee
The Hamiltonian $h$ from \eqref{h} corresponds to the case of the 
isotropic XY chain without external magnetic field, {\it i.e.} to the case where $\gamma=0$ and $\lambda=0$.
\er

The left and right reservoirs carry the inverse temperatures $\beta_L$ and $\beta_R$, respectively. {\it Pour fixer les id\'ees}, we assume,
w.l.o.g., that they satisfy
\be
0
<\beta_L
\le\beta_R
<\infty.
\ee
Moreover, for later use, we set $\beta:=(\beta_R+\beta_L)/2$ and 
$\delta:=(\beta_R-\beta_L)/2$.

We next specify the state in which the system is prepared initially.
It consists of a KMS state at the corresponding temperature for each
reservoir, and, w.l.o.g., of the chaotic state for the sample. For the definition of fermionic quasifree states, see Appendix \ref{app:qf}.
\bd[Initial state]
\label{def:initial}
The initial state 
$\omega_0\in\mQ(\fA)$ is the quasifree state specified by the density
$S_0\in\mL(\hh^{\oplus 2})$ of the form
\be
\label{S0}
S_0
:= s_{0,-} \oplus s_{0,+},
\ee
where the operators $s_{0,\pm}\in\mL(\hh)$ are defined by
\be
s_{0,\pm}
:=(1+\e^{\pm k_0})^{-1},
\ee
and $k_0\in\mL(\hh\simeq\hh_L\oplus\hh_\mS\oplus\hh_R)$ is given by
\be
k_0
:=\beta_L h_{L}\oplus 0 \oplus \beta_R h_{R}.
\ee
Here, for $\alpha=L,\mS, R$, we used the definitions $\hh_\alpha:=\ell^2(\Z_\alpha)$ and $h_{\alpha}:=i_\alpha^\ast h i_\alpha\in\mL(\hh_\alpha)$,  where 
$i_\alpha: \hh_\alpha\to\hh$ is the natural injection defined, for any 
$f\in\hh_\alpha$,  by
$i_\alpha f(x)
:=f(x)$ if $x\in\Z_\alpha$, and zero otherwise.
\ed
\br
Note that  $S_0\in\mL(\hh^{\oplus 2})$ is
well-defined, and that it satisfies the properties of
a density  given in  Definition \ref{def:density} of Appendix \ref{app:qf}.
\er
\br
The one-particle Hilbert space $\hh$ over $\Z=\Z_L\cup\Z_\mS\cup\Z_R$  decomposes as $\hh\simeq \hh_L\oplus\hh_\mS\oplus\hh_R$. It follows from \eqref{h0} in Definition \ref{def:dynamics} that, w.r.t. this
decomposition, the decoupled Hamiltonian $h_0$ does not couple the
different subsystems to each other, indeed, {\it i.e.} we have 
$h_0=h_{L}\oplus h_{\mS}\oplus h_{R}$.
\er

As discussed in the Introduction, we pick the EFP correlation observable in order to study the effect  of the breaking of translation invariance on nonequilibrium expectation values. This
observable is defined as follows.
\bd[EFP]
\label{def:FF}
Let $x_0\in\Z$ and $n\in\N$. The EFP observable $A_{n}\in\fA$ is defined 
by
\be
A_{n}
:= \prod_{i=1}^{2n} B(F_i),
\ee
where, for all $i\in\N$, the form factors $F_i\in\hh^{\oplus 2}$ are given by
\be
F_{2i-1}
&:=& u^i\oplus u^i\, G_1,\\
F_{2i}
&:=& u^i\oplus u^i\, G_2,
\ee
and the initial form factors $G_1, G_2\in \hh^{\oplus 2}$ 
look like
\be
G_1
:=JG_2
:=[0,\delta_{x_0-1}].
\ee
Moreover, the expectation value 
$\rP:\N\to[0,1]$ of the EFP observable $A_n\in\fA$ in the NESS 
$\omega_\rB\in\mE(\fA)$ constructed in Theorem \ref{thm:ness-2} below is denoted by
\be
\label{Pn}
\rP(n)
:=\omega_\rB(A_n).
\ee
\ed
\br
As for the name EFP, note that  
$A_{n}=\prod_{x=x_0}^{x_0+n-1} a^{}_x a^\ast_x$, and that, with
$B_{n}:=\prod_{x=x_0}^{x_0+n-1} a_x$, we have, for any
state $\omega\in\mE(\fA)$, 
\be
0
\le\omega(A_{n})
=\omega(B^{}_{n}B_{n}^\ast)
\le {\|B_{n}\|}^2
\le \prod_{x=x_0}^{x_0+n-1}{\|\delta_x\|}^2
= 1.
\ee
\er
\br
The analysis of this paper can also be carried out for different form factors. If we choose the initial form factors 
$G_i=:[g_{i,1},g_{i,2}]\in\hh^{\oplus 2}$ for $i=1,2$ 
to be of the completely localized form $g_{i,l}=a_{il}\delta_{x_{il}}$ for $a_{il}\in\C$ and $x_{il}\in\Z$ with $l=1,2$, we  cover the case
$G_1=[-\delta_{-1},\delta_{-1}]$ and $G_2=[\delta_0,\delta_0]$.
This choice describes the prominent spin-spin correlations 
$\sigma_1^{(0)}\sigma_1^{(n)}$, see, for example,  Aschbacher and Barbaroux \cite{AB}.
\er

The following definition from Ruelle \cite{R1} introduces the concept of nonequilibrium steady state (NESS) in the framework of $C^\ast$-dynamical systems. For the situation at hand, the $C^\ast$-dynamical system is given in terms of the magnetic Bogoliubov $\ast$-automorphism group 
$\tau_\rB$ on the selfdual CAR algebra $\fA$.
\bd[NESS]
A NESS associated with the $C^\ast$-dynamical system $(\fA,\tau_\rB)$ 
and the initial state $\omega_0\in\mE(\fA)$ is a weak-$\ast$ limit 
point for $T\to\infty$ of the net
\be
\left\{\frac1T\int_0^T\dd t\,\, \omega_0\circ \tau^t_\rB
\,\,\Big|\,\,T>0\right\}.
\ee
\ed

Next, we define the time dependent correlation matrix of the EFP observable $A_{n}\in\fA$ w.r.t. the initial state $\omega_0\in\mE(\fA)$ and the magnetic dynamics $\tau^t_\rB\in{\rm Aut}(\fA)$.
\bd[Correlation matrix]
\label{def:corr}
Let $F_i\in\hh^{\oplus 2}$ for $i\in\N$ be the form factors of Definition 
\ref{def:FF}. For all $t\in\R$, the  skew-symmetric
correlation matrix $\Omega_{n}(t)\in \C_a^{2n\times 2n}:=\{A\in\C^{2n\times 2n}\,|\, A^\rt=-A\}$ (where $A^\rt $ is the transpose of $A$) is
defined, for all $i,j=1,..., 2n$, by its entries
\be
\label{def:cm}
\Omega_{ij}(t)
:=
\begin{cases}
\omega_0(B^\ast({\rm e}^{{\rm i}t H_\rB}JF_i)
B({\rm e}^{{\rm i}t H_\rB }F_j)),
& \mbox{if\, $i<j$},\\
0,
& \mbox{if\, $i=j$},\\
-\,\Omega_{ji}(t),
& \mbox{if\, $i>j$}.
\end{cases} 
\ee
Moreover, the matrices 
$\Omega^\raa_n(t), \Omega^\rap_n(t), \Omega^\rpa_n(t), 
\Omega^\rpp_n(t)\in\C_a^{2n\times 2n}$ are defined, for $i,j=1,...,2n$
 and $i<j$, by
\be
\label{Maa-0}
\Omega^\raa_{ij}(t)
&:=& \omega_0(B^\ast({\rm e}^{{\rm i}t H_\rB}1_{\rm ac}(H_\rB)JF_i)
B({\rm e}^{{\rm i}t H_\rB}1_{\rm ac}(H_\rB)F_j)),\\
\label{Map-0}
\Omega^\rap_{ij}(t)
&:=& \omega_0(B^\ast({\rm e}^{{\rm i}t H_\rB}1_{\rm ac}(H_\rB)JF_i)
B({\rm e}^{{\rm i}t H_\rB}1_{\rm pp}(H_\rB)F_j)),\\
\label{Mpa-0}
\Omega^\rpa_{ij}(t)
&:=& \omega_0(B^\ast({\rm e}^{{\rm i}t H_\rB}1_{\rm pp}(H_\rB)JF_i)
B({\rm e}^{{\rm i}t H_\rB}1_{\rm ac}(H_\rB)F_j)),\\
\label{Mpp-0}
\Omega^\rpp_{ij}(t)
&:=& \omega_0(B^\ast({\rm e}^{{\rm i}t H_\rB}1_{\rm pp}(H_\rB)JF_i)
B({\rm e}^{{\rm i}t H_\rB}1_{\rm pp}(H_\rB)F_j)),
\ee
and are to be completed as in \eqref{def:cm} for 
$i\ge j$. Here, $1_{\rm ac}(H_\rB), 1_{\rm pp}(H_\rB)\in\mL(\hh^{\oplus 2})$ are the spectral projections onto the absolutely continuous and the pure point subspaces of $H_\rB$, respectively.
\ed 

The contributions which will play a role in the large time limit
are defined as follows.
\bd[Asymptotic correlation matrix]
\label{def:acm}
Let $F_i\in\hh^{\oplus 2}$ for $i\in\N$ be the form factors of Definition \ref{def:FF}. The matrices 
$\Omega^\raa_n, \Omega^\rpp_n\in\C_a^{2n\times 2n}$ are defined, 
for $i,j=1,...,2n$ and $i<j$, by
\be
\label{defMaa}
\Omega^\raa_{ij}
&:=&\omega_0(B^\ast(W(H_0,H_\rB) JF_i)B(W(H_0,H_\rB)F_j)),\\
\label{defMpp}
\Omega^\rpp_{ij}
&:=&\sum_{e\,\in\,\spec_\rpp(H_\rB)}
\omega_0(B^\ast(1_e(H_\rB)JF_i)B(1_e(H_\rB)F_j)),
\ee
and are to be completed as in \eqref{def:cm} for 
$i\ge j$. Here, $1_e(H_\rB)\in\mL(\hh^{\oplus 2})$ denotes the spectral projection onto the eigenspace associated with the eigenvalue $e$ in the set of eigenvalues $\spec_{\rm pp}(H_\rB)$ of $H_\rB$, and the wave operator $W(H_0,H_\rB)\in\mL(\hh^{\oplus 2})$ is defined by
\be
\label{wave-H0HB}
W(H_0,H_\rB)
:=\slim_{t\to\infty} {\rm e}^{-{\rm i}tH_0} 
{\rm e}^{{\rm i}tH_\rB}1_{\rm ac}(H_\rB).
\ee
\ed

The following theorem establishes the existence and uniqueness of the NESS
and yields an expression for the EFP in this NESS. 

From now on, whenever an entry of a skew-symmetric matrix is written down, we always assume that the row index is strictly smaller than the column
index. Moreover, for the definition of the Pfaffian, see Appendix \ref{app:qf}.
\bt[NESS and NESS EFP]
\label{thm:ness-2}
There exists a unique quasifree NESS $\omega_\rB\in\mQ(\fA)$ associated
with the $C^\ast$-dynamical system $(\fA,\tau_\rB)$ and the initial state 
$\omega_0\in\mE(\fA)$ whose density $S_\rB\in\mL(\hh^{\oplus 2})$ has the form
\be
\label{SB}
S_\rB
=W^\ast(H_0,H_\rB) S_0 W(H_0,H_\rB)
+\sum_{e\in\, \spec_\rpp(H_\rB)} 1_e(H_\rB) S_0 1_e(H_\rB).
\ee
Moreover, the expectation value of the EFP observable in this NESS 
is given by
\be
\label{NESS2}
\rP(n)
=\pf(\Omega^\raa_n+\Omega^\rpp_n).
\ee
\et
\bprf
We proceed similarly to the proof of Theorem 3.2 in Aschbacher 
{\it et al.} \cite{AJPP2}. To this end, we note that the expectation value in the quasifree initial state $\omega_0\in\mQ(\fA)$ of the
correlation observable $A_n\in\fA$ propagated in time with the magnetic
dynamics $\tau_\rB^t\in{\rm Aut}(\fA)$ can be written,  for all $t\in\R$,
as the Pfaffian of the correlation matrix  
$\Omega_{n}(t)\in\C_a^{2n\times 2n}$ from Definition \ref{def:corr},
\be
\label{pfaff}
\omega_0(\tau_\rB^t(A_n))
=\pf(\Omega_{n}(t)),
\ee
where we used \eqref{JHB} in Remark \ref{rem:JHB} to commute the antiunitary involution $J$ across the unitary group generated by 
$H_\rB\in\mL(\hh^{\oplus 2})$. In order to treat the argument of the
Pfaffian, we make use of assertion (a) in Theorem \ref{thm:spectral} of
Appendix \ref{app:spec} which states that, for the singular continuous
spectrum, we have
\be
\spec_\rsc(H_\rB)
=\emptyset.
\ee
Hence, injecting $1_\rac(H_\rB)+1_\rpp(H_\rB)=1\in\mL(\hh^{\oplus 2})$ 
to the left of $JF_i$ and  $F_j$ in the correlation matrix entry
\be
\Omega_{ij}(t)
&=&\omega_0(B^\ast({\rm e}^{{\rm i}t H_\rB}JF_i)
B({\rm e}^{{\rm i}t H_\rB }F_j))\nonumber\\
&=&({\rm e}^{{\rm i}t H_\rB}JF_i, S_0 {\rm e}^{{\rm i}t H_\rB }F_j), 
\ee
the correlation matrix can be decomposed as
\be
\label{Mdec}
\Omega_{n}(t)
=\Omega^\raa_n(t)+\Omega^\rap_n(t)+\Omega^\rpa_n(t)+\Omega^\rpp_n(t),
\ee
where the matrices on the r.h.s. of \eqref{Mdec} are given in Definition
\ref{def:corr}. Since the NESS is constructed in the large time limit, we
separately study this limit for all the terms in \eqref{Mdec}. So, using that
the initial state is invariant under the decoupled time evolution, 
{\it i.e.} $[H_0,S_0]=0$, the first term can be written as
\be
\Omega^\raa_{ij}(t)
=(\e^{-\ii t H_0}\e^{\ii t H_\rB}1_\rac(H_\rB)JF_i, 
S_0 \e^{-\ii t H_0}\e^{\ii t H_\rB}1_\rac(H_\rB)F_j).
\ee
Thus, with the help of  the Kato-Rosenblum theorem from  scattering theory for perturbations of trace class type (see, for example,  
Yafaev \cite{Y}), we find 
\be
\lim_{t\to\infty}\Omega^\raa_{n}(t)
=\Omega^\raa_{n},
\ee
where we used that $H_0-H_\rB\in\mL^0(\hh^{\oplus 2})$, and the r.h.s.
is given in Definition \ref{def:acm}. For the second  term on the r.h.s. of 
\eqref{Mdec}, we have the bound 
\be
|\Omega^\rap_{ij}(t)|
\le \|1_\rpp(H_\rB)S_0 \e^{\ii t H_\rB} 1_\rac(H_\rB) JF_i\| 
\|F_j\|.
\ee
Since the pure point spectrum of $H_\rB$ consists of the two simple eigenvalues  $\pm e_\rB$, where $e_\rB$ is  given in assertion (c)
of Theorem \ref{thm:spectral} in Appendix \ref{app:spec}, we have 
$1_\rpp(H_\rB)\in\mL^0(\hh^{\oplus 2})$, and, hence, it follows that
\be
\lim_{t\to\infty}\Omega^\rap_{n}(t)
=0.
\ee
The same holds for $\Omega^\rpa_{n}(t)$, of course. For the last term on the r.h.s. of \eqref{Mdec}, using $1_\rpp(H_\rB)=1_{e_\rB}(H_\rB)+1_{-e_\rB}(H_\rB)$, we get
\be
\label{Oppt}
\Omega^\rpp_{ij}(t)
= \sum_{e,e'\in \{\pm e_\rB\}}
{\rm e}^{-{\rm i}t (e'-e)} 
(1_e(H_\rB)JF_i, S_0 1_{e'}(H_\rB)F_j).
\ee
Moreover, since assertion (c) of Theorem \ref{thm:spectral} also states
that $\ran(1_{e_\rB}(H_\rB))\subset \hh\oplus 0$ and $\ran(1_{-e_\rB}(H_\rB))\subset 0\oplus \hh$, and since the density of the initial state 
$S_0\in\mL(\hh^{\oplus 2})$
has the  block diagonal form given in \eqref{S0} of Definition 
\ref{def:initial}, the terms in \eqref{Oppt} for
different energies vanish, and, hence, the time dependence drops out of
\eqref{Oppt}. This leads to
\be
\Omega^\rpp_{n}(t)
=\Omega^\rpp_{n}
\ee
for all $t\in\R$, where the r.h.s. is given in Definition \ref{def:acm}. Finally,
since the Pfaffian $\pf:\C_{a}^{2n\times 2n}\to\C$ is a continuous mapping, we get
\be
\label{omega}
\rP(n)
&=&
\lim_{T\to\infty}\frac1T \int_0^T\!\dd t\,\,
\omega_0(\tau_\rB^t(A_n))\nonumber\\
&=&\lim_{T\to\infty}\frac1T \int_0^T\!\dd t\,\, 
\pf(\Omega_{n}(t))\nonumber\\
&=&\pf(\Omega^\raa_{n}+\Omega^\rpp_{n}).
\ee
Note that we didn't make use of the specific structure of the form
factors $F_i$. Hence, since the algebra of observables $\fA$ is generated by the operators $B(F)$ for $F\in\hh^{\oplus 2}$, and
since the mapping $\fA\ni A\to \omega_0(\tau_\rB^t(A))\in\C$ is
continuous uniformly in $t\in\R$, the relation \eqref{omega} defines 
the unique NESS $\omega_\rB\in\mQ(\fA)$. The form \eqref{SB} of the density $S_\rB$ follows from \eqref{defMaa} and \eqref{defMpp}. Moreover, due to the completeness of the wave operator and Remark \ref{rem:JHB}, 
$S_\rB$  has the defining properties of a density given in Definition 
\ref{def:density} of Appendix \ref{app:qf}. This is the assertion.
\eprf
\section{NESS correlation asymptotics}
\label{sec:asymptotics}

In order to approach the asymptotic behavior of the NESS EFP 
from Theorem
\ref{thm:ness-2}, we start off by studying more closely the two
pieces of the asymptotic correlation matrix given in Definition 
\ref{def:acm}. For this purpose, besides the position space, we will use the momentum space and the energy space defined 
before Theorem \ref{thm:XYness} of Appendix \ref{app:qf} and
in Definition \ref{def:ftilde} of Appendix \ref{app:wave}, respectively.
\bl[$\rac$-structure]
\label{lem:ac}
The asymptotic correlation matrix 
$\Omega^\raa_n\in\C_{a}^{2n\times 2n}$ has the decomposition
\be
\Omega^\raa_n
=\sum_{\sigma=\pm}\Omega^{\raa,\sigma}_n,
\ee
where the matrices $\Omega^{\raa,\pm}_n\in\C_{a}^{2n\times 2n}$ are defined, for all $i,j=1,...,n$, by
$\Omega^{\raa,\pm}_{2i-1\,2j-1}
:=\Omega^{\raa,\pm}_{2i\,2j}
:=\Omega^{\raa,-}_{2i\,2j-1}
:=\Omega^{\raa,+}_{2i-1\,2j}
=0$,
and the nonvanishing entries are given by
\be
\label{Oaa-1}
\Omega^{\raa,-}_{2i-1\,2j}
&:=& (w_-(h,h_\rB)\delta_{i+x_0-1},s_- w_-(h,h_\rB)\delta_{j+x_0-1}),\\
\label{Oaa-2}
\Omega^{\raa,+}_{2i\,2j-1}
&:=& (w_+(h,h_\rB)\delta_{i+x_0-1},s_+ w_+(h,h_\rB)\delta_{j+x_0-1}).
\ee
Here, $s_\pm\in\mL(\hh)$ are the density components of the translation invariant XY NESS given in Theorem \ref{thm:XYness} of Appendix 
\ref{app:qf}, and the wave operators $w_\pm(h,h_\rB)\in\mL(\hh)$ are
defined by
\be
w_\pm(h,h_\rB)
:=\slim_{t\to\pm\infty}\e^{\ii t h}\e^{-\ii t h_\rB}1_\rac(h_\rB),
\ee
where, from now on, all the spectral projections of $h_\rB$ are denoted as the ones for $H_\rB$ given in the Definitions 
\ref{def:corr} and \ref{def:acm} with $H_\rB$ replaced by $h_\rB$.
\el
\bprf
In order to rewrite the absolutely continuous contribution to the asymptotic correlation matrix from Definition \ref{def:acm}, we 
want to take advantage of 
the fact that the operator
\be
\label{WS0W}
S
=W^\ast(H_0,H) S_0 W(H_0,H)
\ee
is the known density of the translation invariant XY NESS ({\it i.e.} the
NESS for $\kp=0$) given in Theorem \ref{thm:XYness} of Appendix 
\ref{app:qf}. For this purpose, we use the chain rule 
 $W(H_0,H_\rB)=W(H_0,H) W(H,H_\rB)$ which is
permissible since $H-H_0, H_\rB-H\in\mL^0(\mH)$ (the wave operators $W(H_0,H)$, $W(H,H_\rB)\in\mL(\hh^{\oplus 2})$ are defined as in 
\eqref{wave-H0HB} with the appropriate replacements). Hence, the absolutely continuous contribution becomes
\be
\label{Omega-aa}
\Omega^\raa_{ij}
=(W(H,H_\rB)J F_i,S W(H,H_\rB) F_j).
\ee
Using the block diagonal structure of the operators $H, H_\rB, S\in\mL(\hh^{\oplus 2})$ and plugging the explicit form of the form factors
from Definition \ref{def:FF} into \eqref{Omega-aa} leads to the 
assertion.
\eprf

In order to evaluate the nonvanishing entries \eqref{Oaa-1} and 
\eqref{Oaa-2} from Lemma \ref{lem:ac}, we determine the action of the wave operators on completely localized wave functions. The main computations are carried out in  Appendix \ref{app:wave}.
\bp[Wave operators]
Let $x\in\Z$ be any site. Then, in momentum space $\hat\hh=L^2(\T)$, the
wave operators $w_\pm(h,h_\rB)\in\mL(\hh)$ act on the completely
localized wave function $\delta_x\in\hh$ as
\be
\label{wave-momentum}
\hat w_\pm(h,h_\rB)\e_x(k)
=\e_x(k)\mp\ii\kp\,\, 
\frac{\e_{|x|}(\mp|k|)}{\sin(|k|)\pm\ii\kp},
\ee
where we set
$\e_x(k):=\hat\delta_x(k)=\e^{\ii kx}$ for all $x\in\Z$
and for all $k\in(-\pi,\pi]$.
\ep
\bprf
Plugging \eqref{rho00}  into \eqref{wave-diag} in Appendix \ref{app:wave}
and applying  $\tilde\ff:\hat\hh\to\tilde\hh$ from \eqref{ftilde} to 
$\e_x$, the action of the wave operator is expressed in energy space 
$\tilde\hh$ as
\be
\label{wave-energy}
\tilde w_\pm(h,h_\rB)\tilde\delta_x(e)
&=&(2\pi)^{-1/2}(1-e^2)^{-1/4} 
\big([(e+\ii\sqrt{1-e^2})^x,(e-\ii\sqrt{1-e^2})^x]\nonumber\\
&&\hspace{3.7cm}\mp\ii\kp \frac{(e\mp\ii\sqrt{1-e^2})^{|x|}}
{\sqrt{1-e^2}\pm\ii\kp}
[1,1]\big).
\ee
Applying $\tilde\ff^\ast:\tilde\hh\to\hat\hh$ from \eqref{ftilde-ast} in
Appendix \ref{app:wave} to \eqref{wave-energy} yields the assertion.
\eprf
\br
The action \eqref{wave-momentum} relates to the action of the wave operator for the one-center $\delta$-interaction on the continuous
line by replacing $\sin(|k|)$ by $|k|$, see, for example, Albeverio 
{\it et al.} \cite{AGHH}.
\er

We next turn to the pure point contribution.
\bl[$\rpp$-structure]
\label{lem:pp}
The asymptotic correlation matrix $\Omega^\rpp_n\in
\C_{a}^{2n\times 2n}$ has the decomposition
\be
\label{reduction}
\Omega^\rpp_n
=\sum_{\sigma=\pm}\Omega^{\rpp,\sigma}_n,
\ee
where the  matrices $\Omega^{\rpp,\pm}_n\in\C_{a}^{2n\times 2n}$ are defined, for all $i,j=1,...,n$, by
$\Omega^{\rpp,\pm}_{2i-1\,2j-1}
:=\Omega^{\rpp,\pm}_{2i\,2j}
:=\Omega^{\rpp,-}_{2i\,2j-1}
:=\Omega^{\rpp,+}_{2i-1\,2j}
:=0$,
and the nonvanishing entries are given by
\be
\label{Opp-1}
\Omega^{\rpp,-}_{2i-1\,2j}
&:=& (1_{\rpp}(h_\rB)\delta_{i+x_0-1},s_{0,-} 1_{\rpp}(h_\rB)\delta_{j+x_0-1}),\\
\label{Opp-2}
\Omega^{\rpp,+}_{2i\,2j-1}
&:=& (1_{\rpp}(h_\rB)\delta_{i+x_0-1},s_{0,+} 1_{\rpp}(h_\rB)\delta_{j+x_0-1}).
\ee
Here, $s_{0,\pm}\in\mL(\hh)$ are the density components of the initial
state given in Definition \ref{def:initial}.
\el
\bprf
Using the block diagonal structures of $H_\rB, S_0\in\mL(\hh^{\oplus 2})$
and plugging the explicit form of the form factors from Definition 
\ref{def:FF} into \eqref{defMpp} leads to the assertion.
\eprf

In order to evaluate the nonvanishing entries \eqref{Opp-1} and 
\eqref{Opp-2} from Lemma \ref{lem:pp}, we determine the form of the
projections onto the pure point subspaces of the magnetic Hamiltonian. 
A summary of its spectral properties is given in Appendix \ref{app:spec}.
\bl[Pure point projection]
\label{lem:pp-1}
The projection onto the pure point subspace of the magnetic Hamiltonian $h_\rB$ satisfies
\be
\dim(\ran(1_\rpp(h_\rB)))
=1,
\ee
and its range is spanned by an exponentially localized eigenfunction 
$f_\rB\in\hh$ of $h_\rB\in\mL(\hh)$ with eigenvalue $e_\rB>1$.
\el
\bprf
See  Theorem  \ref{thm:spectral} in Appendix \ref{app:spec}.
\eprf

Collecting the properties of the absolutely continuous and the pure point contributions to the asymptotic correlation matrix from Lemma 
\ref{lem:ac} to Lemma \ref{lem:pp-1}, we get the following structural assertion.

\begin{figure}
\centering
\includegraphics[width=5cm,height=5cm]{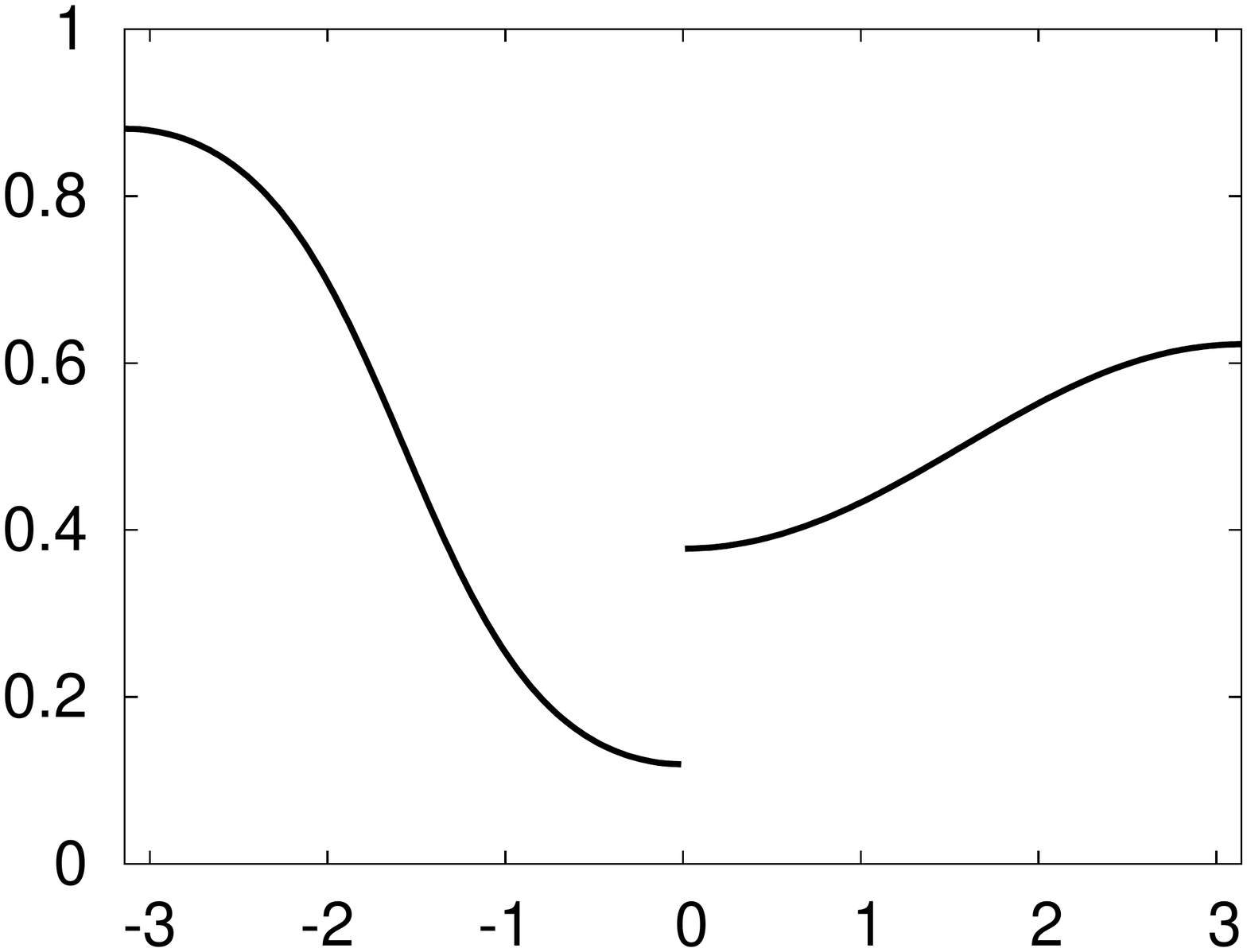}
\includegraphics[width=5cm,height=5cm]{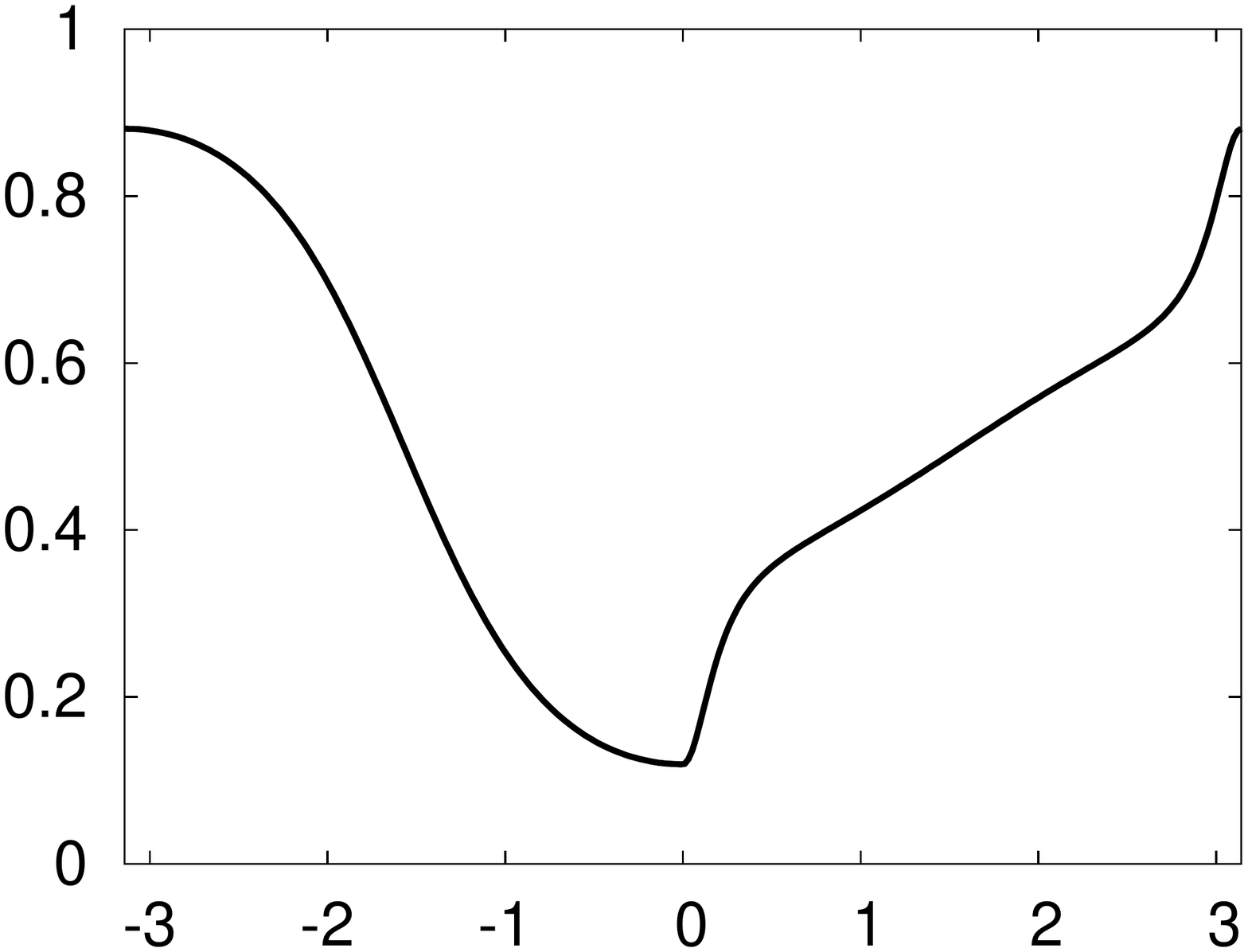}

\caption{The symbol $a(k)>0$ with $k\in(-\pi,\pi]$ for 
$\beta_R=2$, $\beta_L=\tfrac12$, and 
$\kappa=0$ to the left and $\kappa=\tfrac15$ to the right. The nonvanishing magnetic field regularizes the symbol.}
\label{fig:gnu-symbol}
\end{figure}

For the ingredients from Toeplitz theory referred to in the remainder of the present section, see, for example, B\"ottcher and Silbermann 
\cite{BS1,BS2}. Moreover, we denote by $\chi_A:\R\to \{0,1\}$ the characteristic function of the set $A\subset\R$.
\bp[Determinantal structure]
\label{prop:struct}
The NESS EFP is the determinant of the finite section of a Toeplitz  operator, a Hankel operator, and an operator of finite rank. The symbol $a\in L^\infty(\T)$ of the Toeplitz operator reads
\be
\label{a}
a
=\vi_\rB\hat s_{-,L}+(1-\vi_\rB) \hat s_{-,R},
\ee
where the functions $\vi_\rB, \hat s_{\pm,\alpha}\in 
L^\infty(\T)$ with $\alpha=L,R$, are defined, for $k\in(-\pi,\pi]$, by
\be
\label{sLR}
\hat s_{\pm,\alpha}(k)
&:=&\frac12\, (1\pm\tanh[\tfrac12\beta_\alpha\cos (k)]),\\
\label{phiB}
\vi_\rB(k)
&:=&\chi_{[0,\pi]}(k) \, \frac{\sin^2 (k)}{\sin^2 (k)+\kp^2},
\ee
see Figure \ref{fig:gnu-symbol}.  Moreover, the symbol of the Hankel operator is smooth.
\ep
\br
In the limit $\kp\to 0$, we recover the symbol derived in Aschbacher 
\cite{1, 3} for the translation invariant case. 
\er
\br
Note that for nonvanishing coupling, the characteristic function in
\eqref{phiB} is smoothed out. This will play an essential role in the
asymptotic analysis of the corresponding Toeplitz determinant.
\er
\bprf
The total asymptotic correlation matrix, defined by  
$\Omega_{n}:=\Omega^\raa_n+\Omega^\rpp_n\in\C_{a}^{2n\times 2n}$,
has the $2\times 2$ block substructure 
$\Omega_{n}=[A_{ij}]_{i,j=1,...,n}$, where the matrices 
$A_{ij}\in\C^{2\times 2}$ are defined, for $i,j=1,...,n$, by
\be
A_{ij}
:=\begin{cases}
\begin{bmatrix}
0 & b_{ij}\\
c_{ij}& 0
\end{bmatrix}, & \mbox{if  $i<j$},\vspace*{2mm}\\
\begin{bmatrix}
0 & b_{ii}\\
-b_{ii}& 0
\end{bmatrix}, & \mbox{if  $i=j$},\\
-A_{ji}^\rt , & \mbox{if  $i>j$},
\end{cases}
\ee
and the entries are given by
\be
b_{ij}
&:=& \Omega^{\raa,-}_{2i-1\,2j}+\Omega^{\rpp,-}_{2i-1\,2j},
\quad \mbox{if \,} i\le j,\\
c_{ij}
&:=& \Omega^{\raa,+}_{2i\,2j-1}+\Omega^{\rpp,+}_{2i\,2j-1},
\quad \mbox{if \,} i< j.
\ee
In order to rewrite the argument of the Pfaffian in
$\rP(n)=\pf(\Omega_n)$ from Theorem \ref{thm:ness-2}
in a form more suited for the subsequent analysis, we want to apply
a similarity transformation to $\Omega_{n}$. To this end, for any 
$i,j=1,...,2n$ with $i<j$, we denote by $R^{[ij]}\in O(2n)$ the elementary
matrix whose left multiplication with any matrix $A\in\C^{2n\times 2n}$ exchanges the $i$th and $j$th row of $A$ (where $O(n)$ stands for the
orthogonal matrices in $\R^{n\times n}$). Then, using the matrix 
$R\in O(2n)$ defined by 
$R:=\prod_{k=1}^{n-1}\prod_{l=0}^{k-1}R^{[2(n-k)+l,2(n-k)+l+1]}$,
we can transform $\Omega_n$ into off-diagonal block form,
\be
\label{similarity}
R^\rt \Omega_{n}R
=\begin{bmatrix}
 0 & \Theta_n\\
-\Theta_n^\rt  & 0
\end{bmatrix},
\ee
where the matrix $\Theta_n\in\C^{n\times n}$, called the reduced
correlation matrix, is defined by its entries $\Theta_{ij}:=\theta_{ij}$,
and, for all $i,j\in\N$, the numbers $\theta_{ij}$ are given by
\be
\label{Theta-n}
\theta_{ij}
:=\begin{cases}
b_{ij}, & \mbox{if $i\le j$},\\
-c_{ji}, & \mbox{if $i> j$}.
\end{cases}
\ee
Hence, using assertions (a) and (b) from Lemma \ref{lem:pfaffian} of
Appendix \ref{app:qf}, we get
\be
\label{Pn-2}
\rP(n)
&=&
\pf(\Omega_n)\nonumber\\
&=&(-1)^{\frac{n(n-1)}{2}}\pf\bigg(\!\!
\begin{bmatrix}
 0 & \Theta_n\\
-\Theta_n^\rt  & 0
\end{bmatrix}
\!\!\bigg)
\nonumber\\
&=&\det(\Theta_n).
\ee

Let us next analyze the structure of $\Theta_n$. In order to do so, we subdivide the discussion into the following two cases w.r.t. the starting site $x_0\in\Z$ of the EFP string.

{\it Case 1:}\quad $x_0\ge 0$\\
With the help of Lemma \ref{lem:structure} of Appendix \ref{app:matrix}, we make the decomposition $\Theta_n=\Theta_{T,n}+\Theta_{H,n}$, where  the matrix
$\Theta_{T,n}\in\C^{n\times n}$ has the entries
$\Theta_{T,ij}:=\theta_{T,ij}$ given by $\theta_{T,ij}:=b_{T,ij}$ if 
$i\le j$, and $\theta_{T,ij}:=-c_{T,ji}$ if 
$i>j$. Here, for all $i,j\in\N$, we define 
\be
\label{bT}
b_{T,ij}
&:=&(\e_{i-j},\hat s_-\e_0)+(\e_{i-j}, a_-\e_0),
\quad\mbox{if $i\le j$},\\
\label{cT}
-c_{T,ji}
&:=&(\e_{j-i},\hat s_+\e_0)+(\e_{i-j}, a_+\e_0),
\quad\mbox{if $i>j$},
\ee
where $\hat s_\pm, a_\pm\in L^\infty(\T)$ are given in Theorem 
\ref{thm:XYness} of Appendix \ref{app:qf} and Lemma \ref{lem:structure} of Appendix \ref{app:matrix}, respectively.
Similarly, the matrix 
$\Theta_{H,n}\in\C^{n\times n}$ has the entries
$\Theta_{H,ij}:=\theta_{H,ij}$ given by $\theta_{H,ij}:=b_{H,ij}$ if 
$i\le j$, and $\theta_{H,ij}:=-c_{H,ji}$ if 
$i>j$. Here, for all $i,j\in\N$, we define 
\be
\label{bH}
b_{H,ij}
&:=&(e_{i+j},b_-\e_0), 
\quad\mbox{if $i\le j$},\\
\label{cH}
-c_{H,ji}
&:=&(e_{i+j},b_+\e_0), 
\quad \mbox{if $i>j$},
\ee
where $b_\pm\in L^\infty(\T)$ are given in  Lemma \ref{lem:structure}
of Appendix \ref{app:matrix}. 
Using \eqref{sLR} and \eqref{phiB} in  \eqref{bT}, and 
$\hat s_{+,R}-\hat s_{+,L}=-(\hat s_{-,R}-\hat s_{-,L})$ and 
$(\e_{i-j},\hat s_+\e_0)=-(\e_{j-i},\hat s_-\e_0)$ in \eqref{cT}, 
where the last identity is due to Definition \ref{def:density}
of the density in Appendix \ref{app:qf}, we have 
$b_{T,ij}=(\e_{i-j}, a\e_0)$ for $i\le j$ and 
$-c_{T,ji}=(\e_{i-j}, a\e_0)$ for $i>j$. Hence,  $\Theta_{T,n}$  is the
finite section of the Toeplitz operator $T[a]\in\mL(\ell^2(\N))$ generated
by the symbol $a\in L^\infty(\T)$, {\it i.e.}
\be
\Theta_{T,n}
=T_n[a].
\ee
Moreover, as for \eqref{bH} and \eqref{cH}, using \eqref{exp-2} in Remark \ref{rem:FT} of Appendix \ref{app:spec}, we find that 
$b_{H,ij}=(\e_{i+j-1}, b\e_0)$ for $i\le j$ and 
$-c_{H,ji}=(\e_{i+j-1}, b\e_0)$ for $i>j$, where the function 
$b\in C^\infty(\T)$ is defined by
\be
\label{symbol-b}
b(k)
:= \ii\kp\, \frac{\e^{-\ii k(2x_0-1)}}{\sin(k)+\ii\kp}
\left[\frac{(f_B,s_{0,-} f_B)}{e_\rB^2}-\hat s_{-,R}(k)\right].
\ee
Hence, $\Theta_{H,n}$ is the finite section of the Hankel operator 
$H[b]\in\mL(\ell^2(\N))$ generated by the symbol $b\in C^\infty(\T)$, {\it i.e.} 
\be
\Theta_{H,n}
=H_n[b].
\ee
Therefore, it follows from \eqref{Pn-2} that the NESS EFP is the determinant of the finite section of the sum of a Toeplitz and a Hankel operator,
\be
\label{str-1}
\rP(n)
=\det(T_n[a]+H_n[b]),
\ee 
where, in this case, the finite rank operator from the formulation of the assertion
vanishes. 

Let us now turn to the case where the EFP string starts to the left of 
the origin.

{\it Case 2:}\quad $x_0<0$\\
For $n\gg 1+n_0$, where we set $n_0:=-x_0$, we again have from Lemma 
\ref{lem:structure}, that, for all $i,j=1,...,n-n_0$,
\be
\label{x0<0}
\Theta_{n,i+n_0\,j+n_0}
=T_{n-n_0,ij}[a]+H_{n-n_0,ij}[c],
\ee
where we set $c:=\e_{-(2n_0+1)} b$. Defining the operator 
$\Theta: \C^{n_0}\oplus \ell^2(\N)\to \C^{n_0}\oplus\ell^2(\N)$  on $[\xi,f]\in \C^{n_0}\oplus\ell^2(\N)$ by the matrix multiplication with the infinite matrix $\theta_{ij}$ from \eqref{Theta-n}, we have
\be
\label{M}
M:=\Theta-0\oplus (T[a]+H[c])\in\mL^0(\C^{n_0}\oplus \ell^2(\N)).
\ee
Since the reduced correlation matrix satisfies $\Theta_n=R_n 
\Theta R_n \!\!\upharpoonright_{\ran(R_n)}$ with
$R_n:=1\oplus P_n\in\mL(\C^{n_0}\oplus \ell^2(\N))$, it follows from
 \eqref{Pn-2} that
\be
\label{str-2}
\rP(n)
=\det(0\oplus (T_{n-n_0}[a]+H_{n-n_0}[c])+M_n),
\ee 
where $M_n:=R_n M R_n \!\!\upharpoonright_{\ran(R_n)}$.
Hence, we arrive at the assertion.
\eprf

\begin{figure}
\centering
\includegraphics[width=5cm,height=5cm]{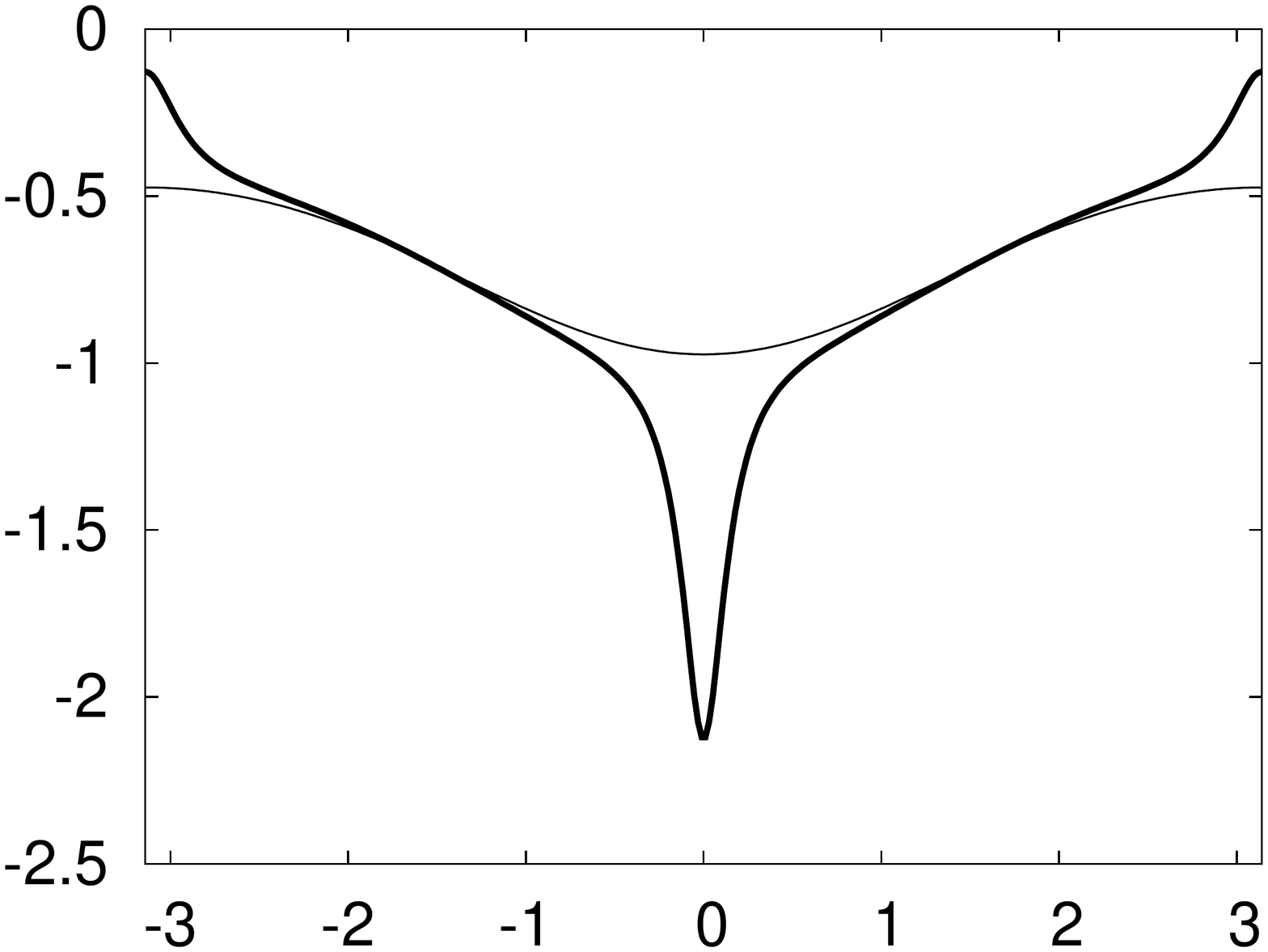}
\includegraphics[width=5cm,height=5cm]{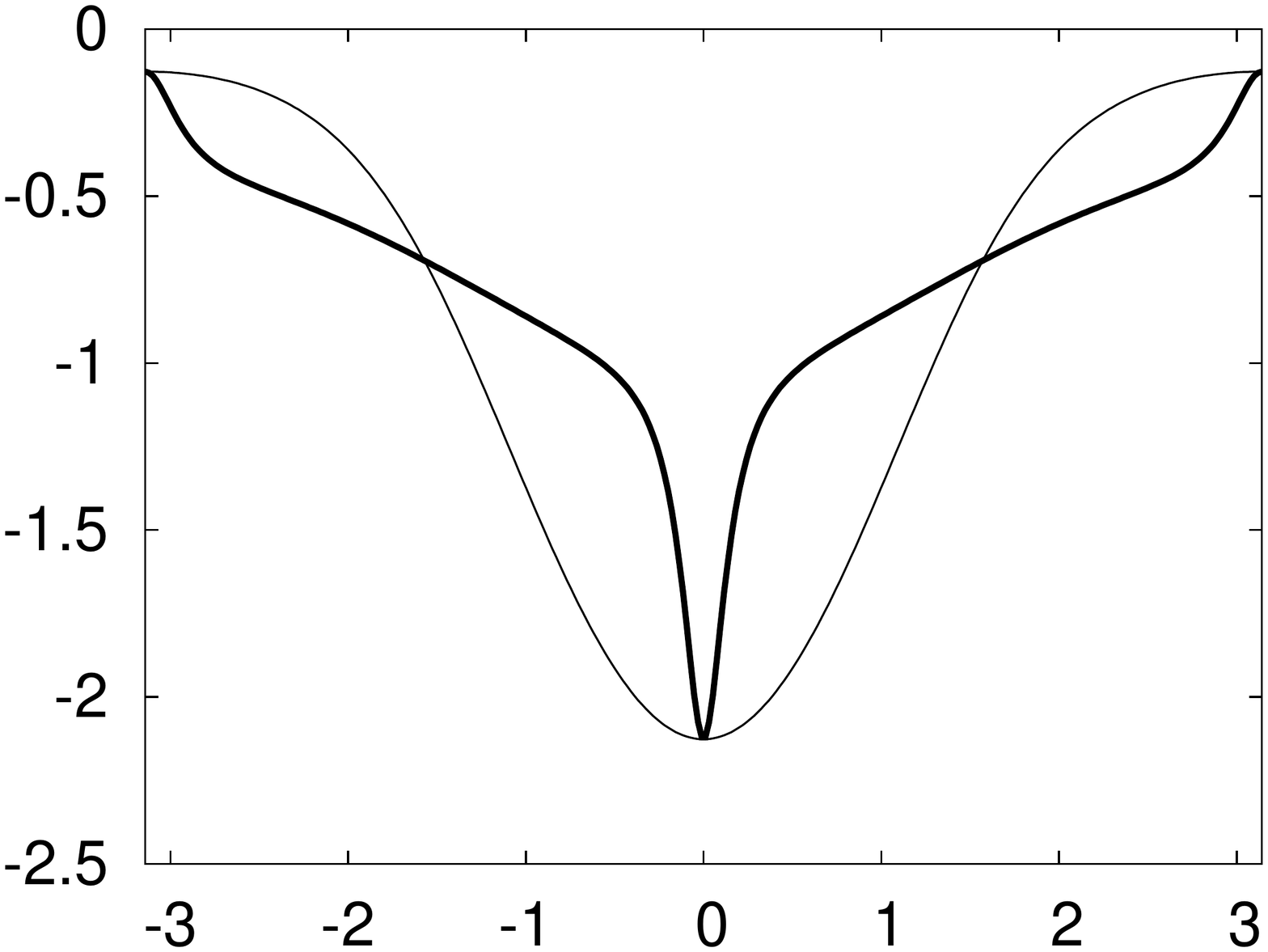}
\caption{For $\beta_R=2$ and $\beta_L=\tfrac12$, the integrand of 
$\Gamma_L$  (left, thin line) and $\Gamma_R$ (right, thin line) compared
to $\Gamma_\rB$  with $\kappa=\tfrac15$ (left and right, thick line).}
\label{fig:gnu-rate}
\end{figure}

We are now ready to formulate our main result on the behavior of the NESS
EFP for large string lengths.
\bt[Exponential decay]
\label{thm:decay}
For $n\to\infty$, the NESS EFP $\rP:\N\to[0,1]$ has an exponentially
 decaying bound,
\be
\rP(n)=\mO(\e^{-\Gamma n}).
\ee
The decay rate $\Gamma:=\Gamma_R+\Gamma_\rB>0$ contains the two parts
\be
\label{rate1}
\Gamma_R
&:=&-\frac12\int_{-\pi}^\pi\frac{\dd k}{2\pi}\,\,
\log [\hat s_{-,R}(k)],\\
\label{rate2}
\Gamma_{\rB}
&:=&-\frac12 \int_{-\pi}^\pi\frac{\dd k}{2\pi}\,\,
\log [\sigma_\rB(k)\hat s_{-,L}(k)
+(1-\sigma_\rB(k))\hat s_{-,R}(k)],
\ee
where the function $\sigma_\rB\in L^\infty(\T)$ is given by
\be
\sigma_\rB(k)
:= \frac{\sin^2(k)}{\sin^2(k)+\kp^2}.
\ee
\et
\br
Note that Theorem \ref{thm:decay} holds for any coupling strength. In the small coupling limit, we recover the exact decay rate from Aschbacher \cite{3}, and the bound derived for general
quasifree systems in Aschbacher \cite{1}.
\er
\br
As can be seen in \eqref{rate2}, the NESS EFP decay rate displays 
a left mover -- right mover structure. It is composed of a left mover
carrying temperature $\beta_R$ and coming from $+\infty$, a left mover
carrying temperature $\beta_R$ having been reflected at the perturbation
at the origin, and a right mover carrying temperature $\beta_L$ having
been transmitted through the origin. This left mover -- right mover structure has already been observed in translation invariant systems
for several types of correlation functions, see  Aschbacher \cite{2, 3} and Aschbacher and Barbaroux \cite{AB}.
\er
\br
Defining $\Gamma_L$ analogously to \eqref{rate1}, we have the rewritings,
for $\alpha=L,R$, 
\be
\label{rate3}
\Gamma_\alpha
&=&-\int_{-\frac\pi2}^{\frac\pi2}\frac{\dd k}{2\pi}\,\,
\log\left[\tfrac14(1-\tanh^2[\tfrac12\beta_\alpha\cos(k)])\right],\\
\Gamma_\rB
\label{rate4}
&=&-\int_{-\frac\pi2}^{\frac\pi2}\frac{\dd k}{2\pi}\,\,
\log\big[\tfrac14(1-[(1-\sigma_\rB(k))\tanh[\tfrac12\beta_R\cos(k)]\nonumber\\
&&\hspace{3.7cm}+\sigma_\rB(k)\tanh[\tfrac12\beta_L\cos(k)]]^2)\big].
\ee 
From \eqref{rate3} and \eqref{rate4}, it immediately follows that, if the system is truly out of equilibrium, {\it i.e.} if $\delta >0$, the decay rates are ordered as
\be
0<
\Gamma_L
< \Gamma_\rB
<\Gamma_R,
\ee
see also Figure \ref{fig:gnu-rate}.
\er
\br
It follows from assertion (a) in Proposition \ref{prop:a-reg} of Appendix
\ref{app:a-reg} that the nonvanishing coupling regularizes the underlying
Toeplitz theory in the sense that the symbol which determines the decay rate is smoother than in the case $\kp=0$, see Figure 
\ref{fig:gnu-symbol}. Namely, the latter case requires Fisher-Hartwig theory and, if $\delta>0$, leads to a strictly positive power law subleading order as given in Aschbacher \cite{3}.
\er
\bprf
Since the Toeplitz symbol $a\in L^\infty(\T)$ from \eqref{a} is 
real-valued, we can make use of the Hartman-Wintner theorem in order to
control the spectrum of the selfadjoint Toeplitz operator 
$T[a]\in\mL(\lN)$. Moreover, due to Proposition \ref{prop:a-reg} of
Appendix \ref{app:a-reg}, we have 
\be
\label{a-cont}
a\in C(\T),
\ee
and, hence,  $\spec(T[a])
=\ran(a)
=[\hat s_{-,R}(0), \hat s_{+,R}(0)]$, where 
$0<\hat s_{-,R}(0)<\hat s_{+,R}(0)<1$ in the temperature range
$0<\beta_L\le \beta_R<\infty$. Therefore, $T[a]\in\mL(\lN)$ is
invertible,
\be
\label{inv}
0\notin \spec(T[a]),
\ee
and the spectrum is independent of the coupling strength. Moreover, since
\eqref{a-cont} and \eqref{inv} hold, the Gohberg-Feldman theorem implies that the sequence $\{T_n[a]\}_{n\in\N}$ is stable,
\be
\label{stable}
\limsup_{n\to\infty} \|T_n^{-1}[a]\|
<\infty.
\ee
For the following analysis, as in the proof of Proposition 
\ref{prop:struct}, we discuss the cases  $x_0\ge 0$ and $x_0<0$ separately. For convenience of exposition, we start with the 
second case.

{\it Case 2:}\quad $x_0< 0$\\
Since we want to analyze the  asymptotic behavior for large $n$ 
with the help of Szeg\H o's strong limit theorem, we write, using 
\eqref{stable},
\be
\label{qt-2}
\frac{\rP(n)}{G(a)^n}
=\frac{\rP(n)}{\det(T_{n-n_0}[a])}
\frac{\det(T_{n-n_0}[a])}{\det(T_{n}[a])}
\frac{\det(T_{n}[a])}{G(a)^n},
\ee
where $G(a)$ is the exponential of the $0$th Fourier coefficient of
$\log(a)$, and  $n_0:=-x_0$ as before. Due to \eqref{str-2} and \eqref{stable}, the first factor on the r.h.s. of \eqref{qt-2} can be
written as
\be
\label{fctr-1}
\frac{\rP(n)}{\det(T_{n-n_0}[a])}
=\det(1+1\oplus T^{-1}_{n-n_0}[a]((-1)\oplus H_{n-n_0}[c]+M_n)).
\ee
Moreover, since we know from Proposition \ref{prop:struct} that 
$c\in C^\infty(\T)$, we also have
\be
\label{b-reg}
c\in L^\infty(\T)\cap B^1_1(\T),
\ee
where $B^\alpha_p(\T)$ are the usual
Besov spaces. Therefore, Peller's theorem allows us to conclude that
($\mL^1(\mH)$ are the trace class operators on the Hilbert space $\mH$)
\be
\label{Hb-L1}
H[b] \in\mL^1(\lN).
\ee
Due to \eqref{M}, \eqref{str-2}, and \eqref{Hb-L1}, 
the r.h.s. of \eqref{fctr-1} converges to the constant 
$K(a,b):=\det(1+1\oplus T^{-1}[a](-1\oplus H[b]+M))$. In order to treat the second factor on the r.h.s. of \eqref{qt-2}, we apply 
Szeg\H o's first limit theorem which is applicable  due to \eqref{a-cont} and \eqref{inv}. Hence,  if we factorize the quotient as 
\be
\label{fctr-2}
\frac{\det(T_{n-n_0}[a])}{\det(T_{n}[a])}
=\prod_{i=1}^{n_0}\frac{\det(T_{n-i}[a])}{\det(T_{n+1-i}[a])},
\ee
each factor on the r.h.s. of \eqref{fctr-2} converges to $1/G(a)$. In order to treat the third factor on the r.h.s. of \eqref{qt-2}, 
we make use of Szeg\H o's strong limit theorem. This theorem states that, since $a(t)>0$ for all $t\in\T$ s.t. $\ind(a)=0$, and since
\be
\label{a-reg2}
a
\in W(\T)\cap B^{1/2}_2(\T),
\ee
which follows from $a\in C^1(\T)\cap PC^\infty(\T)$ in Proposition 
\ref{prop:a-reg} of Appendix \ref{app:a-reg}
($W(\T)$ is the Wiener algebra and $PC^\infty(\T)$ are the piecewise smooth functions), the quotient converges to a
constant usually denoted by $E(a)$. Plugging the foregoing three limits into the r.h.s. of \eqref{qt-2}, we get
\be
\label{lim}
\lim_{n\to\infty}\frac{\rP(n)}{G(a)^n}
=K(a,b)E(a)G(a)^{x_0}.
\ee
Hence, in order to determine an exponential bound on the asymptotic decay, we are left with the computation of the  constant $G(a)$. Decomposing
the integral in the $0$th Fourier coefficient of $\log(a)$ w.r.t. positive and negative momenta and using the fact that $\hat s_{-,\alpha}$ for 
$\alpha=L,R$ and $\sigma_\rB$ are even functions in $k\in(-\pi,\pi]$, we arrive at the expressions \eqref{rate1} and \eqref{rate2} for the decay rate of the bound on the exponential decay.

The case where the EFP string starts at nonnegative sites is simpler
and is treated analogously as follows.

{\it Case 1:}\quad $x_0\ge 0$\\
Writing \eqref{str-1} as in \eqref{qt-2}, where, in this case,  the second factor is absent, we can proceed as for Case 2. In particular,
the determinant of the Toeplitz contribution can again be separated
due to \eqref{stable}, \eqref{Hb-L1} holds for the symbol $b$ satisfying
\eqref{b-reg}, and the constant now reads 
$K(a,c)=\det(1+T^{-1}[a]H[c])$. Finally, the last factor in 
\eqref{lim} is absent. Hence, we arrive at the assertion.
\eprf
\br
The study of the present problem for the anisotropic XY model, {\it i.e.}
for the case where $\gamma\neq 0$ in \eqref{H-XY} of Remark \ref{rem:XY}, is more complicated. 
Not only the Pfaffian structure of the correlation cannot be preserved in the present form, but also one has to cope with Toeplitz theory for operators with nonscalar symbols. We will study this set of problems
for general quasifree systems elsewhere.
\er
\begin{appendix}
\section{Fermionic quasifree states}  
\label{app:qf}  

Let $\fA$ be the selfdual CAR algebra from Definition \ref{def:obs}.
We denote by $\mE(\fA)$ the set of states, {\it i.e.} the normalized
positive linear functionals on the $C^\ast$ algebra $\fA$.
\bd[Density]
\label{def:density}
The density of a state $\omega\in\mE(\fA)$ is defined to be the
operator $S\in\mL(\hh^{\oplus 2})$ with $0\le S^\ast=S\le 1$ and 
$JSJ=1-S$ satisfying, for all $F,G\in \hh^{\oplus 2}$,
\be
\omega(B^\ast(F)B(G))
=(F,SG).
\ee
\ed 

An important class are the quasifree states.
\bd[Quasifree state]
A state $\omega\in\mE(\fA)$ is called quasifree 
if it vanishes on the odd polynomials in the generators and if it
is a Pfaffian on the even polynomials in the generators, {\it i.e.} if, for all 
$F_1,...,F_{2n}\in \hh^{\oplus 2}$ and for any $n\in\N$, we have
\be
\omega(B(F_1)...B(F_{2n}))
=\pf(\Omega_{n}),
\ee
where the skew-symmetric  matrix 
$\Omega_{n}\in\C_{a}^{2n\times 2n}
=\{A\in\C^{2n\times 2n}\,|\, A^\rt =-A\}$ is  defined, for $i,j=1,...,2n$, by
\be
\Omega_{ij}
:=\begin{cases}
\omega(B(F_i)B(F_j)),  & \mbox{if\, $i<j$},\\ 
0,                     & \mbox{if\, $i=j$},\\
-\omega(B(F_j)B(F_i)), & \mbox{if\, $i>j$}.
\end{cases}
\ee
Here, the Pfaffian $\pf: \C_{a}^{2n\times 2n}\to\C$ is given by
\be
\pf(A)
:=\sum_{\pi}\sign(\pi)\prod_{j=1}^nA_{\pi(2j-1),\pi(2j)},
\ee
where the sum is running over all pairings of the set $\{1,2,...,2n\}$,
{\it i.e.} over all the $(2n)!/(2^n n!)$ permutations $\pi$ in the permutation group of $2n$ elements which satisfy $\pi(2j-1)<\pi(2j+1)$ and $\pi(2j-1)<\pi(2j)$, see Figure \ref{fig:pairings}. 
The set of quasifree states is denoted by $\mQ(\fA)$.
\ed

\begin{center}
\begin{figure}
\setlength{\unitlength}{1cm}
\begin{center}
\begin{picture}(22,2)
\multiput(1,0)(0.5,0){6}{\circle*{0.15}}
\put(1.25,0){\oval(0.5,1.5)[t]}
\put(2.25,0){\oval(0.5,1.5)[t]}
\put(3.25,0){\oval(0.5,1.5)[t]}
\put(1.05,-0.6){$\pi=(123456)$}
\put(3.8,0.25){$-$}
\multiput(4.5,0)(0.5,0){6}{\circle*{0.15}}
\put(4.75,0){\oval(0.5,1.5)[t]}
\put(6,0){\oval(1,1.5)[t]}
\put(6.5,0){\oval(1,1.5)[t]}
\put(4.55,-0.6){$\pi=(123546)$}
\put(7.3,0.25){$+$}
\multiput(8,0)(0.5,0){6}{\circle*{0.15}}
\put(8.25,0){\oval(0.5,1.5)[t]}
\put(9.75,0){\oval(1.5,1.5)[t]}
\put(9.75,0){\oval(0.5,1.5)[t]}
\put(8.05,-0.6){$\pi=(123645)$}
\put(10.8,0.25){$-$}
\multiput(11.5,0)(0.5,0){6}{\circle*{0.15}}
\put(12,0){\oval(1,1.5)[t]}
\put(12.5,0){\oval(1,1.5)[t]}
\put(13.75,0){\oval(0.5,1.5)[t]}
\put(11.55,-0.6){$\pi=(132456)$}
\put(14.3,0.25){$+$}
\put(14.8,0.25){\ldots}
\end{picture}
\end{center}
\vspace{0.5cm}
\caption{\label{fig:pairings} Some of the pairings for $n=3$. The total number of intersections  $I$ relates to the signature as 
$\sign(\pi)=(-1)^I$.}
\end{figure}
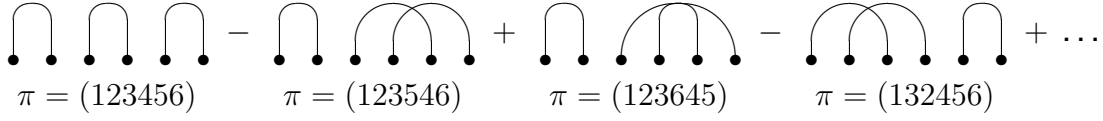
\end{center}

The following lemma has been used in Section \ref{sec:asymptotics}.
\bl[Pfaffian]
The Pfaffian has the following properties.
\begin{enumerate}
\item[(a)] Let $X,Y\in\C^{2n\times 2n}$ with $Y^\rt=-Y$. Then, 
\label{lem:pfaffian}
\be
\pf (XYX^\rt)
=\det(X)\,\pf(Y).
\ee
\item[(b)] Let $X\in\C^{n\times n}$. Then, 
\be
\pf \left(
\begin{bmatrix}
0 &  X\\
-X^\rt & 0
\end{bmatrix}
\right)
=(-1)^{\frac{n(n-1)}{2}}\,\det(X).
\ee
\end{enumerate}
\el
\bprf
See, for example, Stembridge \cite{S}.
\eprf

Next, we state the properties of the NESS for the 
translation invariant case $\kp=0$, the so-called XY NESS. To this end,
let $\ff:\hh\to\hat\hh:=L^2(\T)$ (with unit circle $\T$) be the Fourier transformation defined with the sign convention 
$\hat f(k):=(\ff f) (k):=\sum_{x\in\Z}f(x)\e^{\ii k x}$. Moreover, for any
$a\in\mL(\hh)$, we use the notation $\hat a:=\ff a \ff^\ast$. We then
have the following.
\bt[XY NESS]
\label{thm:XYness}
There exists a unique quasifree NESS $\omega\in \mQ(\fA)$ associated with
the $C^\ast$-dynamical system $(\fA,\tau)$ and the initial state 
$\omega_0\in \mQ(\fA)$ whose density $S\in\mL(\hh^{\oplus 2})$ has the 
form
\be
\label{S}
S
=s_-\oplus s_+,
\ee
where the operators $s_\pm \in\mL(\hh)$ act in momentum space 
$\hat\hh$ as multiplication by 
\be
\label{hat-s}
\hat s_\pm(k)
:=\frac12\,(1\pm \varrho_\pm(k)),
\ee
and the functions $\varrho_\pm:\T\to(-1,1)$ are defined by
\be
\label{rho}
\varrho_\pm(k):=
\tanh\!\big[\tfrac12 (\beta\pm\sign(\sin (k))\delta)\cos (k)\big].
\ee
\et
\bprf
See Aschbacher and Pillet \cite{AP}.
\eprf
\section{Magnetic Hamiltonian}
\label{app:spec}

In this section, we summarize the spectral theory of 
$H_\rB\in\mL(\hh^{\oplus 2})$ needed above. To this end, we denote by 
$\spec_{\rm sc}(A)$, $\spec_{\rm ac}(A)$, and $\spec_{\rm pp}(A)$ the
singular continuous, the absolutely continuous, and the pure point spectrum of the operator $A$, respectively.
\bt[Magnetic spectrum]
\label{thm:spectral}
The magnetic Hamiltonian $H_\rB\in\mL(\hh^{\oplus 2})$ has the following properties.
\begin{enumerate}
\item[(a)] $\spec_{\rm sc}(H_\rB)=\emptyset$

\item[(b)] $\spec_{\rm ac}(H_\rB)=[-1,1]$ 

\item[(c)]
\label{sp-pp}
$\spec_{\rm pp}(H_\rB)=\{\pm e_\rB\}$ with $e_\rB>1$

The eigenvalues $\pm e_\rB$ are simple,  and 
\be
\label{ev-0}
e_\rB
=\sqrt{1+\kp^2}.
\ee
The normalized eigenfunction of $H_\rB$ with eigenvalue $e_\rB$ 
is given by 
$f_\rB\oplus 0\in\hh^{\oplus 2}$, where $f_\rB$ is
exponentially localized, {\it i.e.} for all $x\in\Z$, it has the form
\be
\label{ef-0}
f_\rB(x)
:=\frac{1}{n_\rB}\, \e^{-\lambda_\rB |x|}, 
\ee
and the decay rate and the normalization constant are given by
\be
\label{lambdaB}
\lambda_\rB
&:=&\log(\kp+e_\rB),\\
\label{nB}
n_\rB
&:=&\sqrt{\frac{e_\rB}{\kp}}.
\ee
Moreover, the eigenfunction of $H_\rB$ with eigenvalue $-e_\rB$ reads
 $0\oplus f_\rB\in\hh^{\oplus 2}$.
\end{enumerate}
\et
\bprf
Assertions (a) and (b) are proven 
in a more general context in Hume and Robinson \cite{HR} (which also contains the case of the truly anisotropic XY model without magnetic field, and more general perturbations). The fact that there is no eigenvalue embedded in the continuum
$\spec_{\rm ac}(H_\rB)=\spec_{\rm ac}(H)=\spec(H)=[-1,1]$ also follows
from \cite{HR}. Hence, in order to derive assertion (c), we compute eigenfunctions of the operator $H_\rB$ by looking for solutions of the eigenvalue equation $h_\rB f=ef$ for $e\in\R$ with $|e|>1$ and not identically vanishing $f\in\hh$. Since such $e$ lie in the resolvent set of $h$, we can write $f=-(h-e)^{-1}vf=-\kp f(0) (h-e)^{-1}\delta_0$. By taking the 
scalar product of this equation with $\delta_x$ for any $x\in\Z$, we
have
\be
\label{ef-1}
f(x)
=-\kp f(0) (\delta_x,(h-e)^{-1}\delta_0),
\ee
implying that $f(0)\neq 0$. Plugging $x=0$ into \eqref{ef-1}, we get the 
eigenvalue equation
\be
\label{ev-1}
1+\kp(\delta_0,(h-e)^{-1}\delta_0)
=0.
\ee
Switching to the momentum space
representation and using Cauchy's residue theorem, we get, for all 
$e\in\R$ with $|e|>1$ and all $x\in\Z$, 
\be
\label{ef-2}
(\delta_x,(h-e)^{-1}\delta_0)
=-\sign(e)\frac{(e-\sign(e)\sqrt{e^2-1})^{|x|}}{\sqrt{e^2-1}}. 
\ee
If we plug $x=0$ into \eqref{ef-2} and use the assumption
$\kp>0$, we see that the equation \eqref{ev-1} can be satisfied for
$e>1$ only. Solving \eqref{ev-1} for this case leads to \eqref{ev-0}.
Next, plugging \eqref{ev-0} into \eqref{ef-2}, 
we get $f(x)= f(0)(\kp+e_\rB)^{-|x|}$ from \eqref{ef-1}. Choosing 
$f(0)>0$,  we arrive at \eqref{ef-0} with \eqref{lambdaB} and 
\eqref{nB}.
\eprf
\br
\label{rem:FT}
The Fourier transformation of $f_\rB\in\hh$ is given, for all
$k\in (-\pi,\pi]$, by
\be
\hat f_\rB(k)
=\frac{1}{n_\rB}\,\,
\frac{\kp}{e_\rB-\cos(k)}.
\ee
Cauchy's residue theorem yields that, for $x\in\Z$ with $x\ge 0$,  
we also have 
\be
\label{exp-2}
\e^{-\lambda_\rB x}
=\ii e_\rB\int_{-\pi}^\pi \frac{\dd k}{2\pi}\frac{\e^{-\ii k x}}
{\sin (k) +\ii \kp}.
\ee
The integrand on the r.h.s. of \eqref{exp-2} is used for the extraction of the Hankel symbol in the proof of Proposition \ref{prop:struct}.
\er
\section{Wave operators}
\label{app:wave}

In this section, we use the stationary approach to scattering theory in order to compute the wave operators  $w_\pm(h,h_\rB)\in\mL(\hh)$ appearing in the $\rac$-contribution to the asymptotic correlation matrix from Lemma \ref{lem:ac}. To this end,  we first  need to express the resolvent of the magnetic Hamiltonian by the resolvent of the XY Hamiltonian. This is done in the following lemma.

For any operator $a\in\mL(\hh)$ and any $z\in\C$ in the resolvent set
of $a$, we denote by $r_z(a):=(a-z)^{-1}\in\mL(\hh)$ the resolvent of
$a$ at the point $z$. 
\bl[Magnetic resolvent]
\label{lem:rB}
Let $e\in\R$ and  $\eps>0$. Then, at the points 
$e\pm\ii\eps$, the resolvent
of $h_\rB\in\mL(\hh)$ can be expressed in terms of the resolvent of 
$h\in\mL(\hh)$ as
\be
\label{rB}
r_{e\pm{\rm i}\eps}(h_\rB)
=r_{e\pm{\rm i}\eps}(h)
-\frac{\kp}{1+\kp (\delta_0,r_{e\pm{\rm i}\eps}(h)\delta_0)}\,\, 
(r_{e\mp{\rm i}\eps}(h)\delta_0,\cdot\,)\, r_{e\pm{\rm i}\eps}(h)\delta_0.
\ee
\el
\bprf
In order to simplify notation, we drop the index   $e\pm{\rm i}\eps$ of the resolvents. With the help of the resolvent identity 
$r(h_\rB)=r(h)-\kp r(h)v r(h_\rB)$, we can write, for all $f\in\hh$,
\be
\label{res-1}
r(h_\rB)f+\kp (\delta_0,r(h_\rB)f) r(h)\delta_0
=r(h) f.
\ee
Taking the scalar product of \eqref{res-1} from the left with 
$\delta_0$, we get
\be
\label{res-2}
(1+\kp (\delta_0,r(h)\delta_0)) (\delta_0,r(h_\rB)f)
=(r(h)^\ast\delta_0,f).
\ee
Since, due to 
$(1+\kp (\delta_0,r(h)\delta_0))(1-\kp (\delta_0,r(h_\rB)\delta_0))=1$,
the first factor on the l.h.s. of \eqref{res-2} is nonvanishing, we can solve \eqref{res-2} for $(\delta_0,r(h_\rB)f)$. Plugging the resulting
expression into \eqref{res-1} yields the assertion.
\eprf

In the next definition, we introduce the energy space being the
direct integral decomposition of the absolutely continuous subspace 
of the XY Hamiltonian $h\in\mL(\hh)$ w.r.t. which $h$ is diagonal. 
\bd[Energy space]
\label{def:ftilde}
Let the direct integral over $\spec_\rac(h)$ with fiber $\C^2$ be denoted
by
\be
\tilde\hh
:=L^2([-1,1],\C^2),
\ee
and let us call $\tilde\hh$ the energy space of $h$. Moreover, 
the mapping $\tilde\ff:\hat\hh\to \tilde\hh$ is defined, for all
$\varphi\in\hat\hh$, by
\be
\label{ftilde}
(\tilde\ff \varphi) (e)
:=
(2\pi)^{-1/2}(1-e^2)^{-1/4}\,
[\varphi(\arccos(e)), \varphi(-\arccos(e))].
\ee
We will use the notation 
$\tilde f:=\tilde\ff\ff f$ for all $f\in\hh$, and 
$\tilde a:= \tilde\ff\ff a \ff^\ast\tilde\ff^\ast$ 
for all $a\in\mL(\hh)$, where the Fourier transform 
$\ff:\hh\to\hat\hh$ is defined in Appendix \ref{app:qf}. 
Moreover, the Euclidean scalar product in $\C^2$ will be denoted as 
$\langle\cdot,\cdot\rangle$.
\ed

We then have the following lemma.
\bl[Diagonalization]
The mapping $\tilde \ff\in\mL(\hat\hh,\tilde\hh)$ is unitary, and  the XY Hamiltonian $h\in\mL(\hh)$ acts, on any $\eta\in\tilde\hh$, as the
multiplication by the energy variable $e$,
\be
\label{tilde-h}
(\tilde h\eta) (e)
=e\,\eta(e).
\ee
\el
\bprf
A simple computation shows that ${\tilde\ff}$ is a surjective 
isometry with
${\tilde\ff}^{-1}=\tilde\ff^\ast:\tilde\hh\to\hat\hh$ 
acting on all $\eta=:[\eta_1,\eta_2]\in\tilde\hh$ as
\be
\label{ftilde-ast}
(\tilde\ff^\ast\eta) (k)
=(2\pi)^{1/2}(1-\cos^2(k))^{1/4}\,
[\chi_{[0,\pi]}(k)\,\eta_1(\cos (k))
+\chi_{[-\pi,0]}(k)\,\eta_2(\cos (k))].
\ee
Equality \eqref{tilde-h} then follows  immediately.
\eprf

We introduce the following abbreviations.
\bd[Boundary values]
\label{def:boundary}
Let $e\in\R$ and $\eps>0$. For all $f,g\in\hh$, we define
\be
\label{rho-fg}
\varrho_{f,g}(e\pm\ii\eps)
&:=&(f,r_{e\pm\ii\eps}(h)g),\\
\label{gamma-fg}
\gamma_{f,g}(e,\eps)
&:=&\frac{1}{2\pi\ii}
\left(\varrho_{f,g}(e+\ii\eps)-\varrho_{f,g}(e-\ii\eps)\right).
\ee
Moreover, if the limits exist, we write
\be
\label{rho-0}
\varrho_{f,g}(e\pm\ii 0)
&:=&\lim_{\eps\to 0^+} \varrho_{f,g}(e\pm\ii\eps),\\
\label{gamma-0}
\gamma_{f,g}(e)
&:=&\lim_{\eps\to 0^+} \gamma_{f,g}(e,\eps).
\ee
\ed

The wave operators then have the following form.
\bp[Wave operators]
\label{prop:wave-diag}
In energy space $\tilde\hh$, the action of the wave operators $w_\pm(h,h_\rB)\in\mL(\hh)$ on any $f\in\hh$  has the 
form
\be
\label{wave-diag}
\tilde w_\pm(h,h_\rB)\tilde f(e)
=\tilde f(e)
-\frac{\kp \varrho_{\delta_0,f}(e\pm\ii 0)}
{1+\kp \varrho_{\delta_0,\delta_0}(e\pm\ii 0)}\,\,
\tilde\delta_0(e).
\ee
\ep
\bprf
In order to compute the wave operators $w_\pm(h,h_\rB)\in\mL(\hh)$ with the help of the stationary scheme in scattering theory (see, for example,
Yafaev \cite{Y}), we write them in the weak abelian form
\be
\label{wave0}
w_\pm(h,h_\rB)
=\wlim_{\eps\to 0^+}\, 2\eps \int_0^\infty\dd t\,\,
{\rm e}^{-2\eps t} 1_\rac(h)\e^{\pm \ii th}
\e^{\mp\ii th_\rB}1_\rac(h_\rB).
\ee
Applying Parseval's identity to \eqref{wave0} and using that 
$r_{e\pm\ii\eps}(h)=\pm\ii\int_0^\infty\dd t\,\,
\e^{\mp\ii t(h-(e\pm\ii\eps))}$, we can write, for all $f,g\in\hh$,
\be
\label{wave1}
(f,w_\pm(h,h_\rB)g)
=\lim_{\eps\to 0^+}\frac{\eps}{\pi} \int_{-\infty}^\infty\dd e\,\,
(r_{e\pm\ii\eps}1_{\rm ac}(h)f, 
r_{e\pm\ii\eps}(h_\rB)1_{\rm ac}(h_\rB)g).
\ee
Moreover, if the limits $\eps\to 0^+$ of 
$(r_{e\pm\ii\eps}(h)f, r_{e\pm\ii\eps}(h_\rB)g)$
exist for all $f,g\in\hh$ and almost all $e\in\R$ (the set of
full measure depending on $f$ and $g$), we get
\be
\label{wave2}
(f,w_\pm(h,h_\rB)g)
=\int_{-1}^1\dd e\,\,
\lim_{\eps\to 0^+}\frac{\eps}{\pi} 
(r_{e\pm\ii\eps}(h)f, r_{e\pm\ii\eps}(h_\rB)g)
\ee
because $1_\rac(h)=1$ and $\spec(h)=[-1,1]$. In order to compute the 
limit in \eqref{wave2}, we express the resolvents 
$r_{e\pm\ii\eps}(h)$ of the magnetic Hamiltonian in terms of the 
resolvents $r_{e\pm\ii\eps}(h)$ of the XY Hamiltonian. Plugging 
\eqref{rB} from Lemma \ref{lem:rB} into the scalar product 
on the r.h.s. of \eqref{wave2} and using \eqref{rho-fg} and \eqref{gamma-fg} from Definition \ref{def:boundary}, we have
\be
\label{integrand1}
\frac{\eps}{\pi}\,(r_{e\pm\ii\eps}(h)f,r_{e\pm\ii\eps}(h_\rB)g)
=\gamma_{f,g}(e,\eps)
-\frac{\kp}{1+\kp \varrho_{\delta_0,\delta_0}(e\pm\ii\eps)} \,
\gamma_{f,\delta_0}(e,\eps)\,
\varrho_{\delta_0,g}(e\pm\ii\eps).
\ee
Here, we made use of the fact that, due to the resolvent identity, we have the equality
$\gamma_{f,g}(e,\eps)
=(f,\frac{\eps}{\pi}r_{e\pm\ii\eps}(h)r_{e\mp\ii\eps}(h)g)$. Now, we know that, for any $f,g\in\hh$ and almost all $e\in[-1,1]$, the following limits exist,
\be
\varrho_{f,g}(e\pm\ii 0)
\label{rho-1}
=\pm\pi \ii \frac{\dd  (f,\rho(e)g)}{\dd e}\,
+{\rm p.v.}\!\!\int_{-1}^1 \dd e'\, \frac{1}{e'-e} \,
\frac{\dd (f, \rho(e')g)}{\dd e'},
\ee
where the ${\rm p.v.}$-integral denotes Cauchy's principle value, 
the mapping 
$\rho:\mB(\R)\to\mL(\hh)$ with $\mB(\R)$ the Borel sets on $\R$
is the projection-valued spectral measure
of the XY Hamiltonian $h$, and we used  that
\be
\dd(f, \rho(e)g)
=\chi_{[-1,1]}(e)\, \frac{\dd (f, \rho(e)g)}{\dd e}\,\,  \dd e.
\ee
Moreover, we get from \eqref{gamma-fg} and \eqref{rho-1}, 
\be
\label{gamma-0}
\gamma_{f,g}(e)
=\frac{\dd  (f,\rho(e)g)}{\dd e}.
\ee
Therefore, plugging \eqref{integrand1}, 
\eqref{rho-1}, and \eqref{gamma-0}
 into \eqref{wave2},  we can write
\be
\label{wave3}
(f,w_\pm(h,h_\rB)g)
=(f,g)
-\kp \int_{-1}^1\dd e\,\, 
\frac{\gamma_{f,\delta_0}(e) \varrho_{\delta_0,g}(e\pm\ii 0)}{1+\kp 
\varrho_{\delta_0,\delta_0}(e\pm\ii 0)},
\ee
where, in the first term on the r.h.s., we used 
$\int_{-1}^1\dd e\,\, \gamma_{f,g}(e)=(f,1_{\rm ac}(h)g)=(f,g)$.
In order to write the derivatives in \eqref{gamma-0}
entering \eqref{wave3} more explicitly, we switch to the energy space representation from Definition \ref{def:ftilde}. Using the diagonalization \eqref{tilde-h}, 
we have, for all $f,g\in\hh$, that
\be
\label{derivative}
\frac{\dd  (f,\rho(e)g)}{\dd e}
=\langle \tilde f(e),\tilde g(e)\rangle,
\ee
where we recall from Definition \ref{def:ftilde} that 
$\langle\cdot,\cdot\rangle$ 
denotes the scalar product in the fiber $\C^2$ of the direct integral 
$\tilde\hh=L^2([-1,1],\C^2)$, and $\tilde f=\tilde\ff\ff f$ for all 
$f\in\hh$. Hence, plugging  \eqref{gamma-0} and 
\eqref{derivative} into \eqref{wave3}, we arrive at the assertion.
\eprf

Finally, since the wave operators  $w_\pm(h,h_\rB)\in\mL(\hh)$ appearing in the $\rac$-contribution to the asymptotic correlation matrix act on completely localized wave functions 
$\delta_x\in\hh$ with $x\in\Z$, we compute the terms
$\varrho_{\delta_0,\delta_x}(e\pm\ii 0)$ on the r.h.s. of 
\eqref{wave-diag} in Proposition \ref{prop:wave-diag}.
\bl[Boundary values]
\label{lem:res}
Let $x\in\Z$ and $e\in(-1,1)$. Then, we have
\be
\label{rho00}
\varrho_{\delta_0,\delta_x}(e\pm\ii 0)
=\pm\ii\, \frac{(e\mp\ii\sqrt{1-e^2})^{|x|}}{\sqrt{1-e^2}}.
\ee
\el
\bprf
Let $x\in\Z$ with $x\ge 0$, $e\in (-1,1)$, and $\eps>0$ sufficiently small. Writing $\varrho_{\delta_0,\delta_x}(e-\ii \eps)$ in the momentum
space representation, using Cauchy's residue theorem, and taking the limit $\eps\to 0^+$, we get the expression \eqref{rho00} for
$\varrho_{\delta_0,\delta_x}(e-\ii 0)$. Moreover, using \eqref{rho-fg},
the translation and parity invariance of $h$, {\it i.e.} $[h,u]=0$ and 
$[h,\theta]=0$, respectively, where $\theta:\hh\to\hh$ is defined, for all
$f\in\hh$, by $(\theta f)(x):=f(-x)$, we have, for all $x\in\Z$ with 
$x\ge 0$, that 
\be
\varrho_{\delta_0,\delta_x}(e+\ii \eps)
&=&\overline{\varrho_{\delta_0,\delta_{-x}}(e-\ii \eps)},\\
\varrho_{\delta_0,\delta_{-x}}(e-\ii \eps)
&=&\varrho_{\delta_0,\delta_{x}}(e-\ii \eps).
\ee
This yields the assertion.
\eprf
\section{Asymptotic correlation matrix}
\label{app:matrix}

In this section, we compute the nonvanishing entries of the total
asymptotic correlation matrix used above.
\bl[Structure]
\label{lem:structure}
Let $x_0\ge 0$ and $i,j\ge 1$, or  $x_0<0$ and $i,j\ge 1-x_0$. Then,  the entries of the asymptotic correlation matrix have the structure
\be
\label{O-1}
\hspace{-5mm}\Omega^{\raa,-}_{2i-1\,2j}+\Omega^{\rpp,-}_{2i-1\,2j}
&=&(\e_{i-j},\hat s_-\e_0) 
+(\e_{i-j},a_-\e_0)
+(\e_{i+j}, b_-\e_0),
\quad\mbox{if $i\le j$},\\
\label{O-2}
\hspace{-5mm}\Omega^{\raa,+}_{2j\,2i-1}+\Omega^{\rpp,+}_{2j\,2i-1}
&=&  (\e_{j-i},\hat s_+\e_0)
+(\e_{i-j},a_+\e_0)
+(\e_{i+j}, b_+\e_0),
\quad\mbox{if $i>j$},
\ee
where the functions $a_\pm, b_\pm\in L^\infty(\T)$ are defined by
\be
\label{apm}
a_\pm(k)
&:=&\kp^2 \chi_{[0,\pi]}(k)\frac{\hat s_{\pm,R}(k)-\hat s_{\pm,L}(k)}{\sin^2 (k)+\kp^2},\\
\label{bpm}
b_\pm(k)
&=&(-\ii\kp)\frac{\hat s_{\pm,R}(k)}{\sin (k)+\ii\kp}\, 
\e^{-2\ii k(x_0-1)}
+\frac{\kp}{n_\rB^2} (f_\rB,s_{0,\pm} f_\rB)\, 
\frac{\e^{-2\lambda_\rB(x_0-1)}}{e_\rB-\cos (k)}.
\ee
\el
\bprf
For $i\le j$, plugging the wave operator \eqref{wave-momentum} into 
$\rac$-contribution \eqref{Oaa-1} yields
\be
\label{Om-1}
\Omega^{\raa,-}_{2i-1\,2j}
&=&(\delta_{i+x_0-1},s_-\delta_{j+x_0-1})\nonumber\\
\label{Om-2}
&&+\ii\kp\int_{-\pi}^\pi\frac{\dd k}{2\pi}\,\,\hat s_-(k)\,
\frac{\e^{-\ii(k(i+x_0-1)-|k||j+x_0-1|)}}{\sin(|k|)-\ii\kp}\nonumber\\
\label{Om-3}
&&-\ii\kp\int_{-\pi}^\pi\frac{\dd k}{2\pi}\,\,\hat s_-(k)\,
\frac{\e^{-\ii(|k||i+x_0-1|-k(j+x_0-1))}}{\sin(|k|)+\ii\kp}\nonumber\\
\label{Om-4}
&&+\kp^2\int_{-\pi}^\pi\frac{\dd k}{2\pi}\,\,\hat s_-(k)\,
\frac{\e^{-\ii|k|(|i+x_0-1|-|j+x_0-1|)}}{\sin^2(k)+\kp^2}.
\ee
Moreover, using Lemma \ref{lem:pp-1}, we have 
\be
\Omega^{\rpp,-}_{2i-1\,2j}
&=&\frac{1}{n_\rB^2}\,(f_\rB, s_{0,-} f_\rB)\, 
\e^{-\lambda_\rB (|i+x_0-1|+|j+x_0-1|)}.
\ee
The expressions for $i>j$ are analogous. Then, if $x_0\ge 0$ and 
$i,j\ge 1$, or  $x_0<0$ and $i,j\ge 1-x_0$, we use the translation invariance of $s_\pm$, resolve the absolute values, and  decompose the integrals w.r.t. the sign of the momentum in order to get rid of the 
sign function in the density $\hat s_\pm$. This leads to \eqref{O-1} and \eqref{O-2}.
\eprf
\section{Toeplitz symbol regularity}
\label{app:a-reg}

The following proposition is used in Theorem \ref{thm:decay}.
\bp[Regularity]
\label{prop:a-reg}
The Toeplitz symbol $a\in L^\infty(\T)$ of Proposition 
\ref{prop:struct}  has the following properties.
\begin{enumerate}
\item[(a)] $a\in C^1(\T)\cap PC^\infty(\T)$
\item[(b)] The left and right derivatives
$D_\pm a'(k)$ exist for all $k\in(-\pi,\pi]$, but, for $k_+:=0$ and 
$k_-:=\pi$, we have 
\be
D_+a'(k_\pm)-D_-a'(k_\pm)
=\pm \frac{1}{\kp^2} \frac{\sinh[\tfrac12(\beta_R-\beta_L)]}
{\cosh[\tfrac12\beta_R]\cosh[\tfrac12\beta_L]}.
\ee
\end{enumerate}
\ep
\bprf
From the very form of the symbol given in 
\eqref{a} -- \eqref{phiB}, we get $a\in PC^\infty(\T)$ with jumps at
$k_\pm$, and since $\vi_B\in C(\T)$ for nonvanishing coupling, we also have $a\in C(\T)$. Moreover, the one-sided limits yield  $a'(k_\pm+0)=a'(k_\pm-0)=0$ which is assertion (a), and analogously for assertion (b).
\eprf
\end{appendix}

\noindent\textbf{Acknowledgements}\quad We would like to thank the editor and the referee for their constructive remarks. Moreover, the support provided by the German Research Foundation (DFG) is gratefully acknowledged.


\begin{thebibliography}{99}  

\bibitem{AF} Abanov G A and Franchini F 2003 
Emptiness formation probability for the anisotropic XY spin chain in a magnetic field 
{\it Phys. Lett. A} {\bf 316} 342--9

\bibitem{AGHH} Albeverio S, Gesztesy F, H\o egh-Krohn R, and Holden H
2000 
{\it Solvable models in quantum mechanics}
(AMS: Providence)

\bibitem{A1} Araki H 1968 
On the diagonalization of a bilinear Hamiltonian by a Bogoliubov transformation 
{\it Publ. RIMS Kyoto Univ.} {\bf 6} 385--442
  
\bibitem{A2} Araki H 1971 
On quasifree states of CAR and Bogoliubov automorphisms 
{\it Publ. RIMS Kyoto Univ.} {\bf 6} 385--442

\bibitem{A3} Araki H 1984 
On the XY-model on two-sided infinite chain 
{\it Publ. RIMS Kyoto Univ.} {\bf  20} 277--96  

\bibitem{AH} Araki H and Ho T G 2000 
Asymptotic time evolution of a partitioned infinite two-sided isotropic XY-chain 
{\it Proc. Steklov Inst. Math.} {\bf  228} 191--204  
  
\bibitem{1} Aschbacher W H 2007 
On the emptiness formation probability in quasi-free states
{\it Cont. Math.} {\bf 447}  1--16

\bibitem{2} Aschbacher W H 2007 
Non-zero entropy density in the XY chain out of equilibrium
{\it Lett. Math. Phys.} {\bf 79} 1--16

\bibitem{3} Aschbacher W H 2010
A remark on the subleading order in the asymptotics of the nonequilibrium
emptiness formation probability
{\it Confluentes Math.} {\bf 2} 293-311 (arXiv:1009.1584) 

\bibitem{AB} Aschbacher W H and Barbaroux J M 2007 
Exponential spatial decay of spin-spin correlations in translation invariant quasifree states
{\it J. Math. Phys.} {\bf 48} 113302-1 -- 14

\bibitem{AP}Aschbacher W H and Pillet C A 2003 
Non-equilibrium steady states of the XY chain 
{\it J. Stat. Phys.} {\bf  112} 1153--75

\bibitem{AJPP1} Aschbacher W H, Jak\v si\'c V, Pautrat Y, and Pillet C A 2006 
Topics in non-equilibrium quantum statistical mechanics 
{\it Lecture Notes in Mathematics} {\bf 1882} (New York: Springer) 1--66

\bibitem{AJPP2} Aschbacher W H, Jak\v si\'c V, Pautrat Y, and Pillet C A 2007
Transport properties of quasi-free fermions
{\it J. Math. Phys.} {\bf 48} 032101-1--28

\bibitem{BS1} B\"ottcher A and Silbermann B 1999 
{\it Introduction to large truncated Toeplitz matrices} 
(New York: Springer)

\bibitem{BS2} B\"ottcher A and Silbermann B 2006 
{\it Analysis of Toeplitz operators} 
(Berlin: Springer)

\bibitem{CSP} Culvahouse J W, Schinke D P, and Pfortmiller L G 1969
Spin-spin interaction constants from the hyperfine structure of coupled 
ions
{\it Phys. Rev.} {\bf 177} 454--64

\bibitem{DAT} D'Iorio M, Armstrong R L, and Taylor D R 1983
Longitudinal and transverse spin dynamics of a one-dimensional XY 
system studied by chlorine nuclear relaxation in ${\rm PrCl_3}$
{\it Phys. Rev. B} {\bf 27} 1664--73

\bibitem{FA} Franchini F and Abanov A G 2005 
Asymptotics of Toeplitz determinants and the emptiness formation probability for the XY spin chain 
{\it J. Phys. A: Math. Gen.} {\bf 38} 5069--95

\bibitem{HR} Hume L and Robinson D W 1986
Return to equilibrium in the XY model
{\it J. Stat. Phys.} {\bf 44} 829--48

\bibitem{LSM} Lieb E, Schultz T, and Mattis D 1961 
Two soluble models of an antiferromagnetic chain 
{\it  Ann. Physics}  {\bf 16} 407--66

\bibitem{O} Ogata Y 2002
Diffusion of the magnetization profile in the XX model
{\it Phys. Rev. E} {\bf 66} 066123-1--7

\bibitem{R1} Ruelle D 2000
Natural nonequilibrium states in quantum statistical mechanics
{\it J. Stat. Phys.} {\bf 98} 57 -- 75 

\bibitem{STN} Shiroishi M, Takahashi M, and Nishiyama Y 2001 
Emptiness formation probability for the one-dimensional isotropic XY model 
{\it J. Phys. Soc. Jap.} {\bf 70} 3535--43

\bibitem{SGOVR} Sologubenko A V, Giann\`o K, Ott H R, Vietkine A, and
Revcolevschi A 2001 
Heat transport by lattice and spin excitations in the spin-chain compounds SrCuO$_2$ and Sr$_2$CuO$_3$ 
{\it Phys. Rev. B} {\bf 64} 054412-1--11

\bibitem{S} Stembridge J R 1990
Nonintersecting paths, Pfaffians, and plane partitions
{\it Adv. Math.} {\bf 83} 96 -- 131

\bibitem{W} Widom H 1971 
The strong Szeg\H o limit theorem for circular arcs 
{\it Ind. Univ. Math. J.} {\bf 21} 277--83

\bibitem{Y} Yafaev D R 1992 
{\it Mathematical scattering theory: general theory}
(AMS: Providence, RI)

\end{thebibliography}
\end{document}